\documentclass[]{emulateapj}

\usepackage{verbatim}
\usepackage{multirow}
\bibpunct[, ]{(}{)}{;}{a}{}{,}
\newcommand{\bey}{\begin{eqnarray}}
\newcommand{\eey}{\end{eqnarray}}
\newcommand{\beq}{\begin{equation}}
\newcommand{\eeq}{\end{equation}}
\newcommand{\grb}{GRB~070802}

\shorttitle{Detection of Milky Way type dust}
\shortauthors{El{\'i}asd{\'o}ttir et al.}
\begin{document}

\submitted{Matches version accepted by the ApJ.} 

\title{Dust Extinction in High z Galaxies with GRB Afterglow Spectroscopy - The 2175 {\AA} Feature at  $z=2.45$
\thanks{Based on observations 
collected under progs. ID 079.D-0429(B) and 177.D-0591(P,Q), using the
FORS2 instrument installed at the Cassegrain focus of the Very Large Telescope
(VLT), Unit 1 - Antu, operated by the European Southern Observatory (ESO)
on Cerro Paranal, Chile.}}

\author{
\'A. El{\'i}asd{\'o}ttir\altaffilmark{2,3},
J.~P.~U. Fynbo\altaffilmark{3},
J. Hjorth\altaffilmark{3},
C. Ledoux\altaffilmark{4},
D.~J. Watson\altaffilmark{3},
A.~C. Andersen\altaffilmark{3},
D. Malesani\altaffilmark{3},
P.~M. Vreeswijk\altaffilmark{3},
J.~X. Prochaska\altaffilmark{5},
J. Sollerman\altaffilmark{3,6},
A.~O. Jaunsen\altaffilmark{7}
}

\altaffiltext{2}{Department of Astrophysical Sciences, Princeton University, Princeton, NJ 08544-1001, USA; ardis@astro.princeton.edu }
\altaffiltext{3}{Dark Cosmology Centre, Niels Bohr Institute, University of
Copenhagen, Juliane Maries Vej 30, DK-2100 Copenhagen \O, Denmark}
\altaffiltext{4}{European Southern Observatory, Avenida Alonso de C{\'o}rdova 3107, Casilla 19001, Vitacura, Santiago, Chile}
\altaffiltext{5}{Department of Astronomy and Astrophysics, 
  UCO/Lick Observatory; University of California, 1156 High Street, Santa Cruz, 
  CA 95064; xavier@ucolick.org}
\altaffiltext{6}{Stockholm Observatory, Department of Astronomy, Stockholm University, AlbaNova University Center,  106 91, Stockholm, Sweden}
\altaffiltext{7}{Institute of Theoretical Astrophysics, University of Oslo, PO Box 1029 Blindern, N-0315 Oslo, Norway}

\begin{abstract}
We report the clear detection of the 2175~\AA~dust absorption feature in the optical afterglow spectrum of the
gamma-ray burst (GRB) GRB\,070802 at a redshift of $z=2.45$. This is the
highest redshift for a detected  2175~\AA~dust bump to date,  and it is the first clear
detection of the 2175~\AA~bump in a GRB host galaxy, while several tens
of optical afterglow spectra without the bump have been recorded in the past
decade.  
The derived extinction curve gives $A_V=0.8$--$1.5$ depending on the assumed intrinsic slope.  Of the three local extinction laws, an LMC type extinction gives the best fit to the extinction curve of the host of \grb.   
Besides the 2175~\AA~bump we
find that the spectrum of GRB\,070802 is characterized by unusually strong
low-ionization metal lines and possibly a high metallicity for a
GRB sightline ([Si/H$]=-0.46\pm 0.38$, [Zn/H$]=-0.50\pm 0.68$).  In particular, the spectrum of \grb\ is unique for a GRB spectrum in that it shows clear \ion{C}{1} absorption features, leading us to propose a correlation between the presence of the bump and  \ion{C}{1}.  The gas to dust ratio for the host galaxy is found to be significantly lower than that of other GRB hosts with N(\ion{H}{1})$/A_V=(2.4\pm1.0)\times$~$10^{21}$~cm$^{-2}$~mag$^{-1}$, which lies between typical MW and LMC values.  Our results are in agreement with the tentative conclusion reached by
\citet{gordon2003} that the shape of the extinction curve, in particular the
presence of the bump, is affected by the UV flux density in the environment of
the dust.
\end{abstract}

\section{Introduction}

Dust extinction curves quantify as a function of
wavelength the amount of light `lost' due to scattering and absorption of the light by dust particles
along the line of sight from an object to the observer. In the Milky Way (MW), the
extinction curve has been extensively mapped and has been shown to follow an
empirical single-parameter function for almost all lines of sight \citep{cardelli1989}.
The most characteristic feature of this function is a broad `bump' (i.e.,
excess extinction) centered at $2175$~\AA\ first discovered by \citet{stecher}.
Today, more than 40 years after its discovery, the origin of the feature remains unknown although several candidates have been suggested (see \S~\ref{sec:2175}).

As outlined below (in \S~\ref{sec:grb_after}), a promising method for studying extinction curves
in distant galaxies is to use the spectra of the afterglows of gamma-ray bursts
(GRBs) as the backlight against which the extinction curve can be inferred. The
intrinsic spectral energy distributions (SEDs) of GRB afterglows are very
simple, consisting of a single or a broken power-law from the X-ray band to the
infrared (IR) and hence it is relatively easy to infer the shape of the
extinction curve by measuring the curvature of the optical spectrum, the
deviation of the afterglow spectrum extending from the X-ray extrapolation, or from the
X-ray--IR interpolation. 

In this paper we present the detection of the $2175$~\AA~bump in a GRB
absorption system at $z=2.45$. The detection was briefly reported by 
\citet{2007Msngr.130...43F}, but this paper gives the detailed analysis of our
detection. The detection of the bump has also subsequently been confirmed based
on photometry alone \citep{kruhler08}.  This is the first time that a clear signature of the 2175\ \AA\ feature has been observed in a GRB host galaxy  and the highest redshift detection of the bump to date. 

The paper is organized as
follows: In the remainder of this section we discuss the nature and previous
detections of the 2175~{\AA} feature (\S~\ref{sec:2175}) and the use of GRBs to
infer extinction curves (\S~\ref{sec:grb_after}) in more detail. In \S~\ref{sec:obs} we present
the observations and data analysis. We discuss the properties of the derived
extinction curve in \S~\ref{sec:ext}. The results and possible tracers of the
bump are presented in \S~\ref{sec:dis} and finally we summarize our conclusions
in \S~\ref{sec:con}.

\subsection{Extragalactic extinction curves}
\label{sec:ext_ext}
The study of extinction curves outside the Milky Way is a challenging task and has mostly been limited to galaxies in the Local Group.  While the M31 extinction curve resembles that of the Milky Way \citep{bianchi}, the extinction curves of the Large and Small Magellanic Clouds (LMC and SMC) show significant variation.  Lines of sight in the LMC can be broadly put into two categories, one having extinction curves similar to that of the Milky Way ("LMC average") and the other showing extinction curves with a much less prominent bump and a steeper rise into the UV  \citep["LMC2 supershell"][]{1981MNRAS.196..955N,1999ApJ...515..128M,gordon2003}.  While the phrase "LMC type of extinction" refers to extinction curves of the second type (i.e. less prominent bump and a steep UV rise), it should be kept in mind that Milky Way sightlines also exist in the LMC.  Similarly, for the SMC, the canonical SMC extinction curve shows no evidence for a bump and an even steeper
rise into the UV \citep{1984A&A...132..389P,gordon2003} although one out of five lines of sight shows the 2175\,\AA\ bump \citep{1982A&A...113L..15L}.  It is also worth noting that there are a few known lines of sight in the Milky Way with LMC type of extinction \citep{clayton00} and a single line of sight with a measured SMC type of extinction \citep{2003ApJ...598..369V}.

Relatively little is known about dust extinction outside the Local Group as the method 
of measuring extinction curves by comparing the spectra of single stars is not applicable at
larger distances.
Several methods have been attempted, including studying
statistical samples of reddened and standard quasars \citep{1999ApJ...522..604P,vladilo08}, 
studying individual supernova (SN) Ia lightcurves and using gravitationally lensed quasars.
A recent overview of these studies and their results can be found in 
\citet{eliasdottir2006}.   The SEDs of GRBs may also be used, as discussed in \S~\ref{sec:grb_after}.  
These studies have found varying types of extinction which can deviate significantly from MW-type  extinction, leaving open the question of how common MW type dust is.  
In particular, prior to the present study there 
have only been three robust detections of the
$2175$~\AA~bump in individual extragalactic systems beyond the local group. 
The first is in a lensing galaxy at a redshift of $z=0.83$ \citep{motta2002}. 
The second is from a damped Ly$\alpha$ system at 
$z=0.524$ \citep{junkkarinen2004}.  The third is in an intervening absorber at $z=1.11$ toward GRB~060418 \citep{2006MNRAS.372L..38E}.

Extinction curves are
the prime diagnostic tool available to study dust in the optical/UV regime.
They depend sensitively on the composition of the dust
\citep{henning_etal03}, allowing estimates to be made of the nature and the
origin of dust and its dependence on cosmic time, metallicity, etc.
Furthermore, in studies of distant objects, e.g., using SNe Ia as standard candles or determining flux ratios of strongly lensed 
QSOs, it has become increasingly important to accurately determine the 
amount and wavelength dependence of the dust extinction.

\subsection{The nature and origin of the 2175~{\AA} extinction feature}
\label{sec:2175}
\citet{fitzpatrick+massa86} studied the 2175~{\AA}  interstellar extinction bump in the
direction of 45 reddened stars in the Milky Way, and found that it displays a peak whose central
wavelength $\lambda_{0}$ is remarkably constant with extreme deviation of only
$\sim \pm 17$~{\AA} from the mean position of $\lambda_{0} = 2174.4$~{\AA}. 
As this is significantly larger than the measurement uncertainty it seems
to indicate a real, but small, variation of the peak position.  The full 
width at half maximum (FWHM) of the bump has a larger range of intrinsic 
values, from $360$ to $600$~{\AA}. There seems to be no correlation between 
the width and the central wavelength of the bump.

The fact that the 2175~{\AA} feature becomes progressively fainter in the
MW, LMC and SMC has been attributed to the progressively lower metal 
abundances of the LMC and SMC \citep{fitzpatrick04} 
or to differences in the radiative environment \citep{gordon97,clayton00,mattsson}.   \citet{gordon2003} studied the differences in the extinction of the MW, SMC and LMC finding tentative evidence that the bump strength correlates with dust-to-gas ratio.

\citet{noll07} found in their study of 108 massive, UV-luminous galaxies at 
$1 < z < 2.5$ 
that there is a
correlation between heavy reddening and the presence of a 2175 {\AA} feature. 
The least reddened objects have SEDs consistent with the average featureless and
steep extinction curve of the SMC.   
For their objects at $1 < z < 1.5$ a significant UV bump
is present in galaxies that appear disk-like and which host a rather large 
fraction of intermediate-age
stars (i.e., from 0.2 to 1--2 Gyr old). More clumpy and irregular objects
seem to have extinction curves  
resembling those of nearby starburst galaxies. 

Several candidates have been proposed as the dust particles responsible for
the bump ranging from iron-poor silicate grains \citep{steel+duley} to
carbonaceous materials such as carbon-onions \citep{henrard}, graphite grains
\citep{stecher+donn,draine89} or the polycyclic aromatic hydrocarbons (PAHs)
believed also to be responsible for the prominent broad emission features found in
the mid-infrared \citep{duley+seahra,duley,cecchi-pestellini}. 

Graphite has 
been considered a very promising, though contentious, candidate for
explaining the 2175~{\AA} bump \citep[e.g.,
][]{stecher+donn,fitzpatrick+massa86,mathis94,nicolai90,sorell,draine+malhotra,rouleau,will+aannestad,andersen03,2003ApJ...592..947C}. 
In an extensive investigation,
\citet{draine+malhotra} conclude that if graphite particles are the carriers of
the 2175~{\AA} peak, a variation in their optical properties must be
present, e.g., as a result of varying amounts of impurities, variations in
crystallinity, or changes in its electronic structure due to surface effects.
The observed lack of correlation between the central wavelength and the FWHM of
the peak is therefore a challenge for the hypothesis that graphite
particles are the source of this peak.  

Large PAH molecules are expected to at least contribute to the
2175\,{\AA} feature as the interior carbon atoms have electronic orbitals
closely resembling those in graphene and an oscillator strength for each C
expected to be close to the value for graphite \citep{draine03}. 
\citet{joblin} have suggested that PAHs could be entirely responsible for the 
feature, which has been questioned by \citet{mathis94} due to the lack of an absorption 
feature at about 3000~{\AA} in the spectra of PAHs.  Another argument
against the PAHs as the main contributers is that the constancy in wavelength
along with a rather wide variation in the width, seems surprising if the bump
is caused by a mixture of widely varying materials. Such properties are more likely
to occur as a result of coating (or surface hydrogenation) of carbon grains 
\citep{mathis94,draine+li,cecchi-pestellini}. 

Finally, iron-poor silicates in the form of partially hydrogenated amorphous 
Mg$_{2}$SiO$_{4}$ particles have been suggested as the carriers of the 
2175\,{\AA} peak \citep{steel+duley}.  The absorption is observed in silicates 
of greatly differing composition and under a wide range of different sample
preparations and outgasing conditions.  However,
\citet{mathis96} indicated that the silicates may be problematic as carriers. 

It seems most plausible that the carrier of 2175\,{\AA} peak is a composition of graphitic
and amorphous carbon phases, PAHs and glassy iron poor silicates. Various models have been 
proposed with different compositions \citep[see e.g.,][and references therein]{mathis96, draine03}.

\subsection{GRB afterglows and extinction}
\label{sec:grb_after}
GRB afterglows have very simple intrinsic synchrotron spectra which are 
essentially featureless power laws ($f_\nu \propto \nu^{-\beta}$) across
the wavelength range of interest for measuring extinction curves
\citep[e.g., ][]{1998ApJ...500L..97G,2002ApJ...568..820G,1998ApJ...497L..17S}. This is a significant
advantage over using QSOs or galaxies as background objects. In cases where 
the GRB physics places the cooling frequency between the optical and X-ray 
regime (at the time of observation) there will be a cooling break in the spectrum 
leading to a change in spectral slope of $\Delta\beta=0.5$ (softer spectrum in the 
X-rays). The X-ray slope is usually around 1.0--1.2 \citep[see e.g.,][]{2007ApJ...670..565L} and so the intrinsic 
optical slope is typically between 0.5 and 1.2. Moreover, the optical-NIR range
is typically well-described by a single power-law segment. 

\citet{1998ApJ...495L..99R} first attempted to constrain the extinction 
towards GRB 970508 \citep[see also][]{2001ApJ...553..235R}. Several subsequent 
studies confirmed that a simple SMC-type/linear extinction model provided 
good fits to observations of several GRB afterglows
\citep{1998ApJ...508L..21B,1999ApJ...511L..85C,2001A&A...370..909J,2001A&A...373..796F,2001ApJ...549L...7P,2004ApJ...614..293S, 2004A&A...427..785J,2006ApJ...641..993K}.
Findings of a low dust-to-gas ratios in GRB afterglows exhibiting damped 
Ly$\alpha$ absorption in the GRB host galaxy supports the apparent
prevalence of SMC-type dust in GRB environments
\citep{2001A&A...370..909J,2003ApJ...597..699H,2004A&A...419..927V}.

Gray (wavelength-independent) extinction laws
\citep{2003ApJ...587..135G,2007ApJ...661L...9S,2008ApJ...672..449P,2008MNRAS.386L..87N}
or MW dust \citep{2006ApJ...641..993K,2007MNRAS.377..273S} have been 
suggested in rare instances for GRB environments
although it should be noted that these claims are somewhat model dependent 
(gray dust) or statistically insignificant (MW-type dust).  The 2175\,\AA\ 
feature itself has been searched for and ruled out in 
GRB~050401 \citep{2006ApJ...652.1011W}.
In GRB~000926 \citep{2001A&A...373..796F}
and GRB~020124 \citep{2003ApJ...597..699H}
SMC-type extinction was preferred although the existence of a 2175 \AA\ bump 
was not sampled by the broad-band observations.
\citet{2006A&A...447..145V} found a bump in the spectrum of GRB~991216 but at 
2360 \AA\ if at the redshift of the GRB ($z=1.02$). To be 
consistent with restframe 2175 \AA\ it would have to be due to an even
more distant absorber at $z=1.19$ which was however not detected as 
an absorption system in the afterglow spectrum. The first (and thus far, 
only) clear detection of the 2175 \AA\ bump in the foreground of a GRB, 
consistent with a detected intervening \ion{Mg}{2} absorber ($z=1.1$), was 
reported by \citet{2006MNRAS.372L..38E}.

A significant fraction of GRB afterglows are optically `dark'
\citep{2001A&A...369..373F,2004ApJ...617L..21J,2005ApJ...624..868R}.
It is likely that a large fraction of these afterglows go undetected 
in the optical because of extinction 
\citep[e.g.,][]{2001ApJ...562..654D}.
Recently \citet{2008ApJ...681..453J} and \citet{2008arXiv0803.4100T}
demonstrated the validity of this dust obscuration hypothesis by detecting 
two highly reddened systems. These studies highlight the difficulty in 
probing systems with significant extinction at high redshift: GRBs with 
high extinction are generally not accurately localized (because of the lack 
of an optical afterglow position) and there are no detailed constraints on 
their extinction laws, particularly in the UV, for systems with of order 
$A_V\sim 5$. Such selection effects must be kept in mind before making 
statistical inferences from extinction laws of GRB afterglows.

Studies using the X-ray
extrapolation to set limits on the total extinction
\citep{1999ApJ...523..171V} or using the optical spectral curvature to
determine the reddening \citep{1998Natur.393...43R}
have now been carried out for nearly a decade.  These studies may be further improved if one obtains the largely extinction-free IR flux, typically needing observations
in the mid-IR, and is the subject of our ongoing \emph{Spitzer} campaign
\citep[see also][]{2008arXiv0803.2879H}.

\section{Observations and data analysis}
\label{sec:obs}

\subsection{VLT imaging, spectroscopy and redshift of GRB 070802}

GRB\,070802 \citep{Barthelmy2007GCN} was detected by \textit{Swift} \citep{gehrels2004} on 2007 Aug 2.29682 UT. The ESO VLT started observations about one hour after the burst in Rapid Response Mode, starting with a photometric sequence spanning from the $B$-band through $V$, $R$, and $I$ to the $z$-band in excellent seeing conditions ($\lesssim0.5$ arcsec) and at low airmass ($\lesssim1.4$). We detected the source previously detected by \citet{2007GCN..6695....1B} in all bands (see Table~\ref{tab:photometry}). The source was very red and classifies as a dark burst according to the definition of \citet{2004ApJ...617L..21J}.  Based on 5 R-band points taken from about 1 to 3 hours after the trigger we infer a decay slope of 0.62$\pm$0.05 consistent with the measurement in \citet{kruhler08} at the same time.  We infer a celestial position of $\mathrm{R.A. (J2000)} = 02^{\rm h} 26^{\rm m} 35\fs77$, $\mathrm{decl. (J2000)} = -55\arcdeg 31\arcmin 39\farcs2$ (calibrated against USNO-B1.0).  Following the afterglow detection, spectroscopy was immediately started, using the FORS2 instrument equipped with the grism 300V ($\sim 10$~{\AA} FWHM spectral resolution). A total of 5400~s of exposure were acquired, split into three exposures, with mean epoch of Aug 2.38935 UT (2.2 hours after the burst).  The spectra were obtained in excellent seeing conditions
(about $0\farcs5$) and a relatively small airmass of 1.2. We used a $1\farcs0$ 
slit and therefore slitloss should be neglicible. The spectra were flux calibrated
using a spectrum of the spectrophotometric standard star LTT1020 obtained on the
same night.
\begin{deluxetable}{@{}lcccl@{}}
\tablecaption{Log of photometric observations\label{tab:photolog}}
\tablewidth{0pt}
\tablehead{
Mean epoch & Time since & Exposure & Filter & Mag.\\
(2007 Aug. UT) & trigger (hr) & times (s) & & (Vega)
}
\startdata
2.33589  & 0.94  &  60  & $z$  & 20.39$\pm$0.05 \\
2.33753  & 0.98  &  40  & $I$  & 21.36$\pm$0.10 \\
2.33443  & 0.90  &  30  & $R$  & 21.86$\pm$0.06 \\
2.34353  & 1.12  &  30  & $R$  & 22.03$\pm$0.06 \\
2.34451  & 1.14  &  60  & $R$  & 22.18$\pm$0.06 \\
2.35186  & 1.32  &  60  & $R$  & 22.26$\pm$0.05 \\
2.42306  & 3.03  &  60  & $R$  & 22.72$\pm$0.05 \\
2.33881  & 1.01  &  40  & $V$  & 22.69$\pm$0.15 \\
2.34066  & 1.05  & 150  & $B$  & 23.99$\pm$0.28 \\
\enddata
\label{tab:photometry}
\end{deluxetable}

\begin{figure}
\includegraphics[angle=0,width=\columnwidth,clip=]{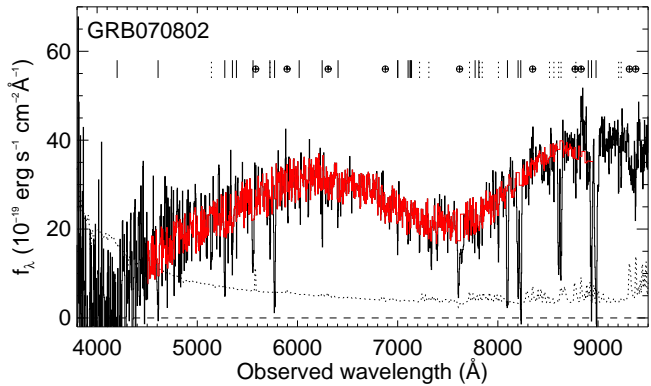}
\caption{The flux-calibrated spectrum of the GRB\,070802 afterglow vs. observed wavelength.
Metal lines at the host redshift are marked with solid lines whereas the lines
from the two intervening systems are marked with dotted lines. The broad
depression centered around 7500 \AA\ is caused by the 2175 \AA\  extinction
bump in the host system at $z_{\rm abs}=2.4549$.  The cleaned spectrum used for
the extinction curve analysis in \S~\ref{sec:ext} is overplotted in red. Telluric
features are indicated with $\earth$.}
\label{fig:spectrum}
\end{figure}

The optical spectrum of the afterglow is shown in Fig.~\ref{fig:spectrum}.
The afterglow spectrum is characterized by a red continuum
with a broad dip centered at $\lambda_{\rm obs}\approx 7500$ \AA. There is
a break in the spectrum at $\lambda_{\rm obs}\approx 4250$ \AA\ blueward of
which little residual flux, if any, is detected. We identify the latter
feature as the imprint of the Ly$\alpha$ forest of absorption lines at
$z_{\rm abs}<2.5$. Numerous narrow absorption lines are detected
throughout the spectrum (see Table~\ref{tab:lines}). The highest redshift absorber 
is a strong metal line system at $z_{\rm abs}=2.4549$, which we adopt as the GRB
host galaxy redshift. In addition we detect two intervening \ion{Mg}{2} absorption
systems at $z=2.078$ and $z=2.292$. The system at $z=2.078$ is very strong
with a restframe equivalent width of the 2796 \AA\  line of $W_r=3.9$~\AA.
\begin{deluxetable}{lllll}
\tablecaption{Absorption lines in the afterglow spectrum of \grb.}
\tablewidth{0pt}
\tablehead{
$\lambda_{\textrm{obs}}$ [\AA] &
$\lambda_{\textrm{rest}}$ [\AA] &
\hspace*{2mm} $z$ &
Feature &
$W_o$ [\AA]
}

\startdata

4200.0 & 1215.7 & 2.455   & Ly$\alpha$ &  \\
4607.9 & 1335.3 & 2.4529 & \mbox{C\,{\sc ii}} / \mbox{C\,{\sc ii}}* & 16$\pm$6 \\
5273.9 & 1526.7 & 2.4544 & \mbox{Si\,{\sc ii}} & 10$\pm$2 \\
5350.9 & 1548.2/1550.8 & 2.4533 & \mbox{C\,{\sc iv}} & 7.3$\pm$1.8 \\
5390.7 & 1560.3 & 2.4549 & \mbox{C\,{\sc i}}\tablenotemark{a} & 2.4$\pm$1.2  \\
5556.4 & 1608.4 & 2.4546 & \mbox{Fe\,{\sc ii}} & 10.5$\pm$1.6 \\
5724.5 & 1656.9 & 2.4550 & \mbox{C\,{\sc i}} & 4.7$\pm$1.1 \\
5771.2 & 1670.7 & 2.4544 & \mbox{Al\,{\sc ii}} & 14.3$\pm$1.2 \\
6246.5 & 1808.8 & 2.4549 & \mbox{Si\,{\sc ii}} & 3.9$\pm$1.0 \\
6404.9 & 1854.7 & 2.4533 & \mbox{Al\,{\sc iii}} & 2.4$\pm$1.0 \\
7000.1 & 2026.1 & 2.4550 & \mbox{Zn\,{\sc ii}} & \multirow{2}{*}{4.2$\pm$1.0} \\
7000.1 & 2026.5 & 2.4553 & \mbox{Mg\,{\sc i}} &  \\
7104.1 & 2056.3 & 2.4548 & \mbox{Cr\,{\sc ii}} & 2.0$\pm$0.9 \\
7124.8 & 2062.2 & 2.4545 & \mbox{Cr\,{\sc ii}} & \multirow{2}{*}{2.53$\pm$0.9} \\
7126.3 & 2062.7 & 2.4545 & \mbox{Zn\,{\sc ii}} & \\
7810.7 & 2260.8 & 2.4549 & \mbox{Fe\,{\sc ii}} & 6.6$\pm$1.0 \\
8094.1 & 2344.2 & 2.4515 & \mbox{Fe\,{\sc ii}} & 16.5$\pm$2.0 \\
8200.5 & 2374.5 & 2.4536 & \mbox{Fe\,{\sc ii}} & 12.7$\pm$1.5 \\
8228.0 & 2382.8 & 2.4531 & \mbox{Fe\,{\sc ii}} & 18.8$\pm$1.5 \\
8277.1 & 2396.4 & 2.4540 & \mbox{Fe\,{\sc ii}}* & 2.9$\pm$1.4 \\
8308.7 & 2405.6 & 2.4539 & \mbox{Fe\,{\sc ii}}** & 2.8$\pm$1.4 \\
8902.8 & 2576.9 & 2.4549 & \mbox{Mn\,{\sc ii}} & 5.0$\pm$1.2 \\
8934.4 & 2586.7 & 2.4538 & \mbox{Fe\,{\sc ii}} & 19.5$\pm$1.3 \\
8979.5 & 2600.2 & 2.4534 & \mbox{Fe\,{\sc ii}} & 29.0$\pm$1.3 \\
8515.5 & 2586.7 & 2.2920 & \mbox{Fe\,{\sc ii}} & 2.3$\pm$0.6 \\
8560.6 & 2600.2 & 2.2923 & \mbox{Fe\,{\sc ii}} & 2.2$\pm$0.6 \\
9208.6 & 2796.3 & 2.2931 & \mbox{Mg\,{\sc ii}} & 1.8$\pm$0.6 \\
9226.5 & 2803.5 & 2.2911 & \mbox{Mg\,{\sc ii}} & 1.4$\pm$0.6 \\
5141.4 & 1670.7 & 2.0774 & \mbox{Al\,{\sc ii}} & 6.0$\pm$1.7 \\
7218.7 & 2344.2 & 2.0794 & \mbox{Fe\,{\sc ii}} & 3.4$\pm$1.0 \\
7311.4 & 2374.5 & 2.0791 & \mbox{Fe\,{\sc ii}} & 4.5$\pm$1.0 \\
7335.7 & 2382.8 & 2.0786 & \mbox{Fe\,{\sc ii}} & 4.9$\pm$1.0 \\
8005.9 & 2600.2 & 2.0790 & \mbox{Fe\,{\sc ii}} & 4.7$\pm$1.1 \\
8607.0 & 2796.3 & 2.0780 & \mbox{Mg\,{\sc ii}} & 11.3$\pm$0.6 \\
8631.0 & 2803.5 & 2.0787 & \mbox{Mg\,{\sc ii}} & 11.5$\pm$0.6 \\
8780.1 & 2853.0 & 2.0775 & \mbox{Mg\,{\sc i}}  & 2.7$\pm$1.1 \\

\enddata

\tablenotetext{a}{The line is blended with \ion{C}{1}$^*$ and \ion{C}{1}$^{**}$}
\tablecomments{The equivalent widths are given in the observer frame, $W_o$.}
\label{tab:lines} 
\end{deluxetable}

\subsubsection{\ion{H}{1} content and metal lines}
\begin{figure}
\includegraphics[angle=0,width=\columnwidth,clip=]{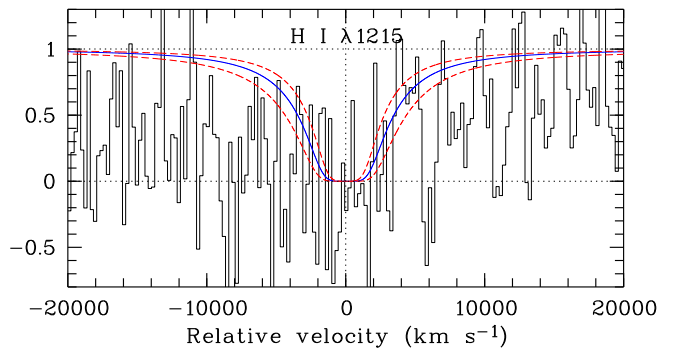}
\caption{A section of the spectrum centered on the position of Ly$\alpha$
at $z_{\rm abs}=2.4549$. Overlaid is the best fitting DLA profile and associated 1~$\sigma$ uncertainty, 
corresponding to $\log{N}$(\ion{H}{1})$=21.5\pm0.2$.}
\label{fig:HI}
\end{figure}

We measured the total neutral hydrogen column density (in units of cm$^{-2}$
throughout) of the absorber at
the GRB redshift by fitting a damped Ly$\alpha$ absorption line
profile fixed at the redshift of the detected metal lines. 
Both the saturated core and the red damped wing of the
Ly$\alpha$ line are constrained by the residual flux at $\lambda_{\rm obs}
\ge 4250$ \AA. 
We derived $\log N($\ion{H}{1}$)=21.5\pm 0.2$ (see Fig.~\ref{fig:HI}).

As listed in Table~\ref{tab:lines},
a large number of metal species are identified at the GRB 
redshift.
The detected lines from singly ionized
elements are very strong compared to other GRB- or QSO-DLA
sightlines. For instance,  $W_r=2.9$~\AA\ for \ion{Si}{2}\,$\lambda1526$ is larger than for any LBG/GRB/QSO-DLA observed to date, hinting at a 
high metallicity \citep{prochaska2008}. The restframe equivalent width of \ion{Si}{2}\,$\lambda1808$, $W_r=1.13$ \AA\, is
also the largest ever detected in any extragalactic
sightline. Assuming the latter line is located on the linear part of
the curve of growth implies that $\log N($\ion{Si}{2}$)=16.3$. This
is a very conservative lower limit to the actual column density and
translates into a conservative lower limit on the metallicity of
[Si/H$]>-0.8$. Note also that Si could be depleted onto dust grains
\citep{petitjean2002}.

\begin{figure}
\includegraphics[angle=0,width=\columnwidth,clip=]{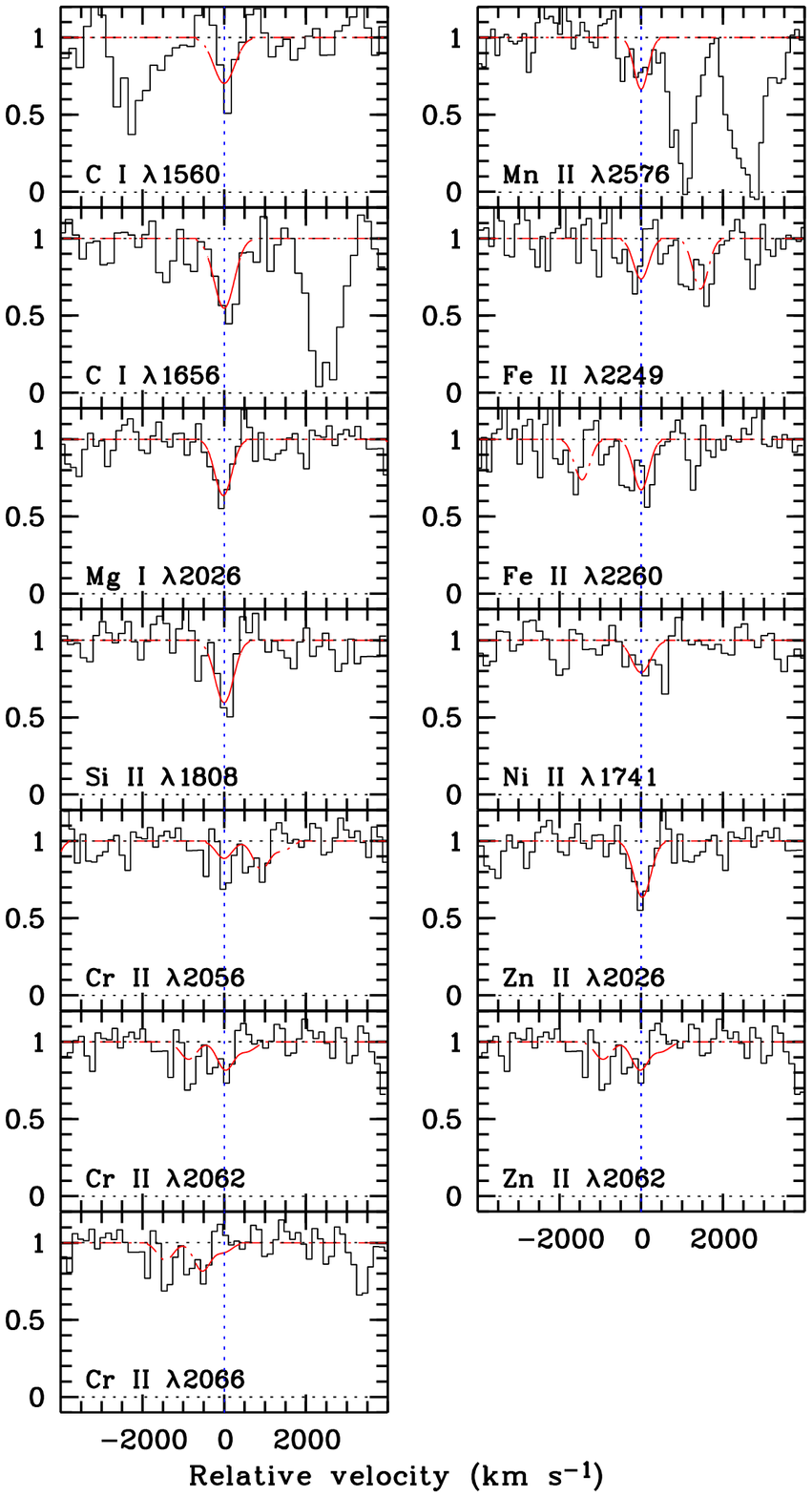}
\caption{Our Voigt-profile fits to low-ionization metal lines.
Best-fitting parameter values are given in Table~\ref{tab:voigt}.}
\label{fig:metals}
\end{figure}

We further performed Voigt-profile fitting (see Table~\ref{tab:voigt}) using
some of the weakest (least saturated) metal lines (see Fig.~\ref{fig:metals}).
The absorption line at $\lambda _{\rm obs}\approx 7000$ \AA\ is likely to be a
blend of \ion{Zn}{2} and \ion{Mg}{1}\,$\lambda$2026 with possibly similar
strengths. It is therefore important to fit both transitions simultaneously.
The other \ion{Zn}{2} ($\lambda$2062) line is also blended (with one of the
\ion{Cr}{2} triplet lines). Therefore, the best line to estimate the
metallicity remains \ion{Si}{2}\,$\lambda$1808. All lines were simultaneously
fitted with a single component constraining their redshifts as well as their
broadening parameter (assumed to be purely turbulent) to have identical values.
We find [Si/H$]=-0.46\pm 0.38$ and also consistently get [Zn/H$]=-0.50\pm 0.68$.
Assuming the \ion{Zn}{2}\,$\lambda$2026 line is located on the linear part of the
curve-of-growth implies that [Zn/H]>-0.35 but the actual metallicity could
be lower than that due to the unknown contribution of the \ion{Mg}{1}\,$\lambda$2026 line
to the observed feature (see above).
\begin{table}
\caption {Ionic column densities of the GRB system.
\label{tab:voigt}}
\begin{center}
\begin{tabular}{lll}
\hline
\hline
Ion & Transition lines used& $\log N$ \\
 & ({\AA})&      (cm$^{-2}$)\\
\hline
\ion{Si}{2} & 1808 & 16.6(0)$\pm$0.3(2)  \\ 
\ion{Zn}{2} & 2026; 2062 & 13.6(7)$\pm$0.6(5)  \\ 
\ion{Mg}{1} & 2026 & 14.4(1)$\pm$0.6(4)  \\ 
\ion{C}{1} & 1560; 1656 & 14.9(5)$\pm$0.2(5)  \\ 
\ion{Fe}{2} & 2249; 2260\tablenotemark{a} & 16.1(6)$\pm$0.1(8)  \\ 
\ion{Cr}{2} & 2056; 2060; 2066\tablenotemark{a} & 14.0(4)$\pm$0.4(0)  \\ 
\ion{Ni}{2} & 1741 & 14.8(9)$\pm$0.2(8)  \\ 
\ion{Mn}{2} & 2576 & 13.8(9)$\pm$0.1(8)  \\ 

\hline
\tablenotetext{a}{Even though the FeII2249 and CrII2066 lines are not individually detected above 2~$\sigma$ in the spectrum we still use the appropriate regions in the spectrum to constrain the fits. }
\end{tabular}
\tablecomments{The broadening parameter was $b=$123$\pm$74~km s$^{-1}$ in all cases.}
\end{center}
\end{table}

The large error bars on the measured column densities are mainly a
consequence of the large uncertainty on the broadening parameter $b$
which we find to be: $b=123\pm 74$ km s$^{-1}$. 
One might argue that $b$ could be significantly smaller
(e.g., $b<30$ km s$^{-1}$) in which case the column densities would be
much larger than found in our best fit. However, it has been shown
that the width of metal line profiles in DLAs is larger at higher
metallicities \citep{2006A&A...457...71L,prochaska2008}. Therefore,
a large $b$ value is consistent with the relatively high
metallicity inferred, which is among the highest measured for
any GRB \citep{2006A&A...451L..47F}.

The measured abundance ratios are consistent with what is seen in 
other GRB- or QSO-DLAs. We find that the mean depletion of Fe-peak
elements (Fe, Cr and Mn) compared to Si is $\sim 0.5$ dex. This is
not particularly high, but yet significant implying that 
dust is present in this
system. According to our best fit, the Si metallicity could be as
large as solar. This in fact is quite likely taking into account that
Si is affected by dust depletion \citep{petitjean2002}.

\subsubsection{Neutral species}
In the afterglow spectrum, we clearly identify neutral carbon lines,
namely \ion{C}{1}\,$\lambda\lambda$1560,1656 (blended with
\ion{C}{1}$^\star$ and \ion{C}{1}$^{\star\star}$;  see
Fig.~\ref{fig:metals}). These are the first ever clear detections of \ion{C}{1}
in a GRB-DLA which typically have very low \ion{C}{1}$/$\ion{C}{2} ratios \citep{2007ApJ...666..267P}. The two features are very strong. We measured
$W_r=1.4$ \AA\ at $\lambda_{\rm obs}\approx 5720$ \AA\ which is dominated
by \ion{C}{1}\,$\lambda$1656 absorption. Using the above fit results,
we get log N$($\ion{C}{1}$)/N($\ion{Si}{2}$)=-1.65$. This can be
compared to QSO-DLAs from Fig. 10 of \citet{srianand2005} showing
that H$_2$ is most likely present as well in this system, although this can not be verified as it falls too much in the blue into  the Ly$\alpha$ forest, and because of low signal to noise and spectral resolution. 
On the other hand, the strength of the detected \ion{C}{4} doublet
line is only modest and similar to most other GRB sight-lines.

\subsubsection{Non detection of \ion{Fe}{2}$^\star$ in \grb.}
Over the past few years, excited lines of \ion{Fe}{2} have been detected along
several GRB sightlines \citep[e.g.,
][]{2004ApJ...614..293S,2005ApJ...634L..25C}, which have been shown to be due
to indirect excitation by UV photons produced by the GRB afterglow
\citep{2006ApJ...648...95P,2006ApJ...648L..89D,2007A&A...468...83V}. For two
specific GRBs, modeling of the afterglow flux exciting a cloud of \ion{Fe}{2}
and/or \ion{Ni}{2} atoms has led to an estimate of the distance between the GRB
and the bulk of the neutral absorption material (responsible for low-ionization
lines from ions such as \ion{C}{1}, \ion{O}{1}, \ion{Fe}{2}, \ion{Cr}{2},
\ion{Mn}{2}, \ion{Si}{2}, \ion{Zn}{2}, \ion{Ni}{2}): ~2 kpc for GRB 060418
(Vreeswijk et al. 2007), and ~0.5 kpc for GRB 050730 (Ledoux et al. 2009, submitted). For the GRB 070802 sightline we detect consistent features at the expected positions of the \ion{Fe}{2}$^*\lambda$2396  and
 \ion{Fe}{2}$^{**}\lambda$2405 lines at z=2.455 but, since the lines are likely to be at the same time
slightly saturated and blended, we cannot infer from them secure column densities.
Based on the distance estimates in previous GRBs we
note that the bulk of the neutral material, which likely corresponds to the
region where the extinction bump is originating, is probably at least 0.5 kpc
away from GRB 070802.

\subsubsection{Constraints on the $z=2.078$ foreground absorber extinction}
\label{sec:foreground}
The foreground absorption system at $z=2.078$ exhibits
strong \ion{Mg}{2} ($W_r^{\lambda 2796}=3.9$ \AA) and \ion{Fe}{2} lines with
equivalent width ratio $W_r^{\lambda 2796}/W_r^{\lambda 2600}\la 2$.
Additionally the detected \ion{Mg}{1}\,$\lambda$2852 line has $W_r=1.0$ \AA. These
characteristics make this absorber very likely a DLA
\citep[see, e.g., ][their fig.~11]{2006ApJ...636..610R}.  We note that such a strong \ion{Mg}{2} absorber is very rare for QSO sightlines, but it has been suggested that GRB sightlines may have stronger absorbers although the reason for this is still unknown \citep{2006ApJ...648L..93P} .

To constrain $A_V$ from the foreground absorber we look at the metal lines as metallicities and dust depletion factors (and hence the dust content) correlate in DLAs \citep{2003MNRAS.346..209L}.
The non-detection of \ion{Zn}{2}\,$\lambda$2026 leads to an upper limit on
the zinc column density.
Unfortunately, the constraint we can derive from the spectrum,
$\log N($\ion{Zn}{2}$)<13.5$ at the $3\sigma$ level, is fairly weak
because there is an absorption feature immediately bluewards of the
expected position of the \ion{Zn}{2} line.
Moreover, we cannot use \ion{Si}{2} instead of \ion{Zn}{2} because the expected
position of the \ion{Si}{2}\,$\lambda$1808 line at $z=2.078$ is blended with
\ion{Fe}{2}\,$\lambda$1608 from the GRB host galaxy absorber at $z=2.455$.
The best constraint on $A_V$ therefore comes from the observed \ion{Fe}{2} lines
at $z=2.078$. The ratio of the weakest detected \ion{Fe}{2} line, i.e.,
\ion{Fe}{2}\,$\lambda$2374, at $z=2.455$ to that at $z=2.078$ is 4.3. Because
of saturation effects, this value is a lower limit on the ratio of the
\ion{Fe}{2} column densities. Altogether, the weaker \ion{Fe}{2} lines
at $z=2.078$ and the non-detection of \ion{Zn}{2}\,$\lambda$2026 at $z=2.078$  are an indication of lower metallicity because in DLAs the dust depletion
factor is only a slowly decreasing function of metallicity \citep[e.g., ][]{2003MNRAS.346..209L,2008A&A...481..327N}.
A factor of $>4.3$ leads to $\log N($\ion{Zn}{2}$)<13$ and therefore $A_V<0.25$
\citep[][their fig.~1]{2005A&A...444..461V}.

An alternative way to estimate the contribution of the foreground absorber is to us the relation reported by \citet[][their eq. (18)]{menard2008} which uses the equivalent width of the \ion{Mg}{2} line to estimate $E(B-V)$.  Assuming a standard MW extinction law, with $R_V=3.1$, this translates into $A(V)\approx 0.06$ which is well below the limit derived above.

\subsection{X-ray observations}
The X-ray telescope onboard \emph{Swift} observed the afterglow of
GRB\,070802 beginning observations in photon counting mode at 147\,s
after the trigger time ($t_0 =$ 07:07:26 on 2 August 2007). The data
were extracted and reduced in a standard way using the HEAsoft
software (version 6.2) and the most recent calibration files. The
X-ray spectrum (with a total exposure time of 3419\,s in the interval
$t_0+147$--$t_0+16800$\,s) was extracted  and fit with a power-law with
absorption fixed at the Galactic level ($2.9\times10^{20}$\,cm$^{-2}$) \citep{1990ARA&A..28..215D} 
and freely variable absorption at the redshift of the host galaxy. The
best fit resulted in a power-law with a photon spectral index 
$\Gamma = 2.02^{+0.17}_{-0.15}$ ($\beta = 1.02^{+0.17}_{-0.15}$, $68\%$ confidence level) 
and an equivalent
absorbing hydrogen column density
of $9.7_{-0.4}^{+0.6}\times10^{21}$\,cm$^{-2}$ at $z=2.45$ assuming solar abundances. To
produce the X-ray segment of the near-infrared to X-ray spectral
energy distribution (SED), this X-ray spectrum was normalized to the
flux level obtained from the X-ray lightcurve at the SED time
($t_0+3517$\,s) and corrected for both absorptions. No significant evidence
(above the $1\,\sigma$ level) was found for spectral variations over the period
of the spectrum, by comparing the first 400\,s and the remainder of the observation.

\subsection{The host galaxy}
\begin{figure}
\includegraphics[angle=0,width=\columnwidth,clip=]{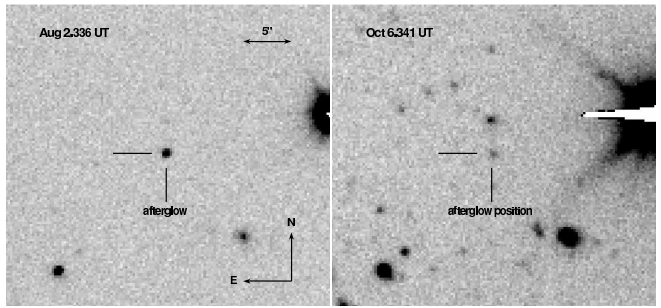}
\caption{The field of \grb. The left panel shows the afterglow as detected about 1~hr after the burst. The right panel shows a deep late-time exposure of the same field (65 days after the burst), taken when the afterglow had faded away, and revealing a host galaxy at the GRB position. Each panel is $30\arcsec \times 30\arcsec$ wide.}
\label{fig:finding}
\end{figure}

$K$- and $R$-band observations of the host galaxy of GRB 070802 were obtained with the ESO VLT during the nights of 27 September 2007 and 5 October 2007  
(57 and 65 days after the GRB). At the afterglow position, a source is clearly 
detected with $R = 25.03 \pm 0.10$ (Fig.~\ref{fig:finding}).
A faint source is also visible in $K$, with magnitude $K = 21.70 \pm 0.25$. 
The centroid of the host is offset by $0.15 \pm 0.04$ arcsec with respect to 
the afterglow position (1.2~kpc at $z = 2.455$). From the observed $K$ 
magnitude (rest frame 6270~\AA), and assuming a power-law spectrum consistent 
with the $R-K$ color ($f_\nu \propto \nu^{-1.3}$), we can estimate the 
absolute $B$-band magnitude of the host to be $M_B = -21.0$ (uncorrected 
for extinction) which is fairly luminous for a GRB host \citep{2006Natur.441..463F,2008arXiv0803.2718S}.

\section{The extinction curve}
\label{sec:ext}
The spectrum shows strong extinction features, with a clear detection of a bump
near $\lambda=2175$~\AA\ in the restframe of the host (Fig.~\ref{fig:spectrum}).  To extract the continuum part of the spectrum to use for the extinction curve fitting we start by selecting the wavelength range $4500$--$9000$\AA.   We then remove emission and absorption lines by recursively fitting the spectrum to an 8th order polynomial removing parts which deviate by more than $2.2\sigma$ (see Fig.~\ref{fig:spectrum}).

The intrinsic flux of the afterglow is modeled as
\begin{equation}
f_\nu^{\mathrm{intr}}=f_{\nu,0} \left(\frac{\nu}{\nu_0}\right)^{-\beta}
\end{equation}
(where $\nu$ is frequency, $\beta$ the intrinsic slope and $f_{\nu,0}$ is the flux at $\nu_0$) but is partially extinguished due to dust along the line of sight.
The observed flux is therefore given by
\begin{eqnarray}
f_\nu^{\mathrm{obs}}&=&f_\nu^{\mathrm{intr}} 10^{-0.4A_\lambda^{\mathrm{total}}}\\
    &= &f_{\nu,0}  \left(\frac{\nu}{\nu_0}\right)^{-\beta} 10^{-0.4\left(A_\lambda+A_\lambda^{\mathrm{Gal}}\right)}
\end{eqnarray}
where the total extinction is $A_\lambda^{\mathrm{total}}=A_\lambda+A_\lambda^{\mathrm{Gal}}$, 
$A_\lambda$ is the extragalactic extinction along the line of sight as a function of the wavelength $\lambda$ and $A_\lambda^{\mathrm{Gal}}$ is the extinction in the Milky Way.  The extragalactic extinction can be caused by both extinction in the host of the GRB and in foreground objects, i.e. $A_\lambda=A_\lambda^{\mathrm{host}}+A_\lambda^{\mathrm{fore}}$.  The observed flux is corrected for extinction in the Milky Way using the maps of \citet[][$E(B-V)=0.025$, $R_V=3.1$]{schlegel1998}, leaving
\begin{equation}
f_\nu=f_{\nu,0}  \left(\frac{\nu}{\nu_0}\right)^{-\beta} 10^{-0.4 A_\lambda}
\end{equation}
which is the intrinsic flux of the burst affected only by extragalactic extinction along the line of sight.   Solving for the extinction, we find
\begin{eqnarray}
A_\lambda&=&-2.5\log_{10}\left(\frac{f_\nu}{f_{\nu,0}}\left(\frac{\nu}{\nu_0}\right)^\beta\right)\\
&=&-2.5\log_{10}\left(\frac{f_\nu}{f_{\nu,0}}\left(\frac{x}{x_{0}}\right)^\beta\right)
\end{eqnarray}
where $x\equiv1/\lambda=\nu/c$ is the wavenumber.
The absolute extinction along the line of sight can therefore be determined if the underlying spectral slope $\beta$ and the normalization $f_{\nu,0}$ at $x_0$ are known.  

\begin{figure}
\includegraphics[angle=270,width=\columnwidth,clip=]{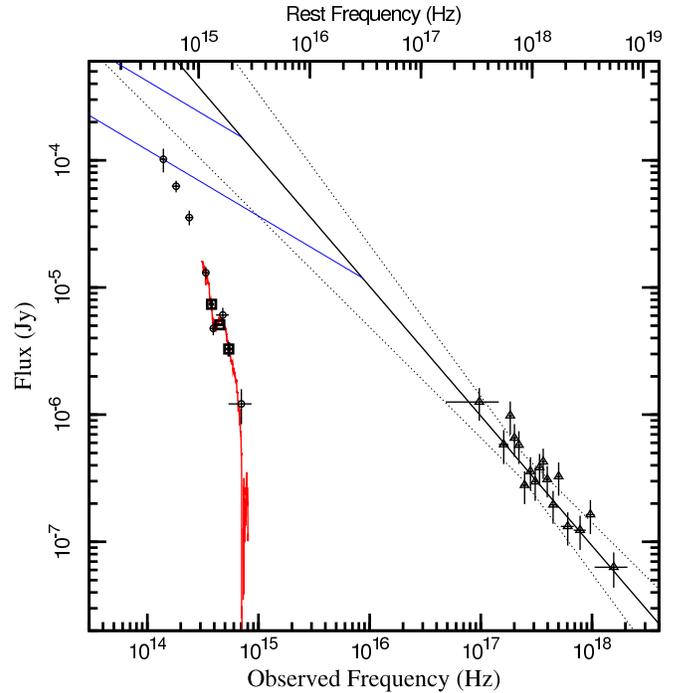}
\caption{Spectral energy distribution of the afterglow of \grb\ at $t_0+3517$~s.  The plot shows X-ray data (triangles, corrected to the time frame of the optical photometric points) and the derived intrinsic powerlaw spectrum (black solid line) and the 68\% confidence levels (black dashed lines).  The spectrum (red curve) is scaled to the $R$-band.  The squares are the photometric points from this paper, the diamonds are the photometric points of \citet{kruhler08}.  Assuming that the intrinsic slope of the GRB can be described as a single powerlaw (i.e. $\beta=\beta_X=1.02_{-0.15}^{+0.17}$), the difference in the observed spectrum and the extrapolated X-ray spectrum is interpreted as being due to extinction.  The blue lines correspond to the upper and lower limits of the intrinsic power law given a cooling break in the spectrum.  The intrinsic slope in the optical is then $\beta=\beta_X-0.5=0.52_{-0.15}^{+0.17}$ and the normalization is determined by the location of the cooling break, which can be anywhere between the optical and X-ray data sets with the additional constraint that the optical data points can not be brighter than the intrinsic curve.  The intrinsic powerlaw spectrum with a cooling break corresponding to the 68\% confidence levels have been omitted for clarity.}
\label{fig:xraynobreak}
\end{figure}

The simple shape of the intrinsic spectra of GRB afterglows (see \S~\ref{sec:grb_after}) allows us to constrain both $\beta$ and the normalization.  Assuming that there is no cooling break in the intrinsic spectrum, one may use the slope and normalization derived from the X-ray data, i.e. $\beta=\beta_X$.  Alternatively, assuming that there is a cooling break in the spectrum, the slope must be $\beta=\beta_X-0.5$ while the normalization depends on the location of the cooling break.  These two possibilities will both be addressed here and are depicted in Fig.~\ref{fig:xraynobreak}.  In addition, we will start by assuming that all the extinction is caused by dust in the host galaxy of \grb, i.e. that 
\begin{equation}
A_\lambda=A_\lambda^{\mathrm{host}}.
\end{equation}
The possible contribution from the foreground \ion{Mg}{2} absorption systems (\S~\ref{sec:foreground}) will be addressed in \S~\ref{sec:ext_fore}. 

We fit the derived extinction curves to four types of extinction laws, i.e. MW type extinction as parametrized by \citet{cardelli1989}, LMC and SMC type extinction as parametrized by \citet{pei1992} and the modified parametrization for the UV of \citet[][FM]{fitzpatrick2007}.  A description of these extinction laws and how the \citet{pei1992} parametrization compares to the \citet{gordon2003} SMC and LMC extinction curves is given in Appendix~\ref{app:ext}.  The parametrization of \citet{fitzpatrick2007} does not directly tie the slope to the bump, but keeps the parameters of each independent.  It therefore has much greater freedom in fitting the curve and is able to accurately trace the derived extinction.  The parametrization however assumes that $A_V$ and $R_V$ have been independently derived from the optical extinction.  As our data do not reach into the restframe optical, we have opted to fix $R_V=3.1$ and allow $A_V$ to vary while fitting the UV.   We also tested choosing $R_V=2.7$ and $R_V=3.5$ and find that it leads to a change in $A_V$ and $c_1$ (one of the parameters of the FM extinction law, see App.~\ref{app:ext}) of around $10\%$ but does not significantly affect the other parameters. 

We compare the extrapolation of our fits to the infrared photometric points of \citet{kruhler08}.  We used the spectral energy distribution (SED) from GROND created by
citet{kruhler08} to derive the near-infrared data we show here simply by
extrapolating the mean SED time of \citet{kruhler08} to our mean SED time using
our estimate of the optical decay rate.  We note that these points are not used in constraining the fits.

\subsection{The extinction assuming no cooling break}
If there is no cooling break between the X-ray and optical regimes,
clearly the X-ray and optical emission lie on the same power-law.
Under this assumption, the intrinsic slope and normalisation in the
optical can therefore be derived from a simple extrapolation of the
X-ray data.  Any discrepancy between the observed optical flux and the
extrapolation is then interpreted as extinction along the line of
sight. A slight complication is introduced by the non-simultaneity of
the observations in different bands, since the flux decays with time.
However the uncertainty introduced by this correction is not large,
since the optical photometry mean times are $I=t_0+3517$\,s,
$V=t_0+3627$\,s and $R$ is interpolated between observations at
$t_0+3249$\,s and $t_0=3949$\,s, and the X-ray flux is derived from a
lightcurve that covers this time period. 

\begin{figure}
\includegraphics[angle=0,width=\columnwidth,clip=]{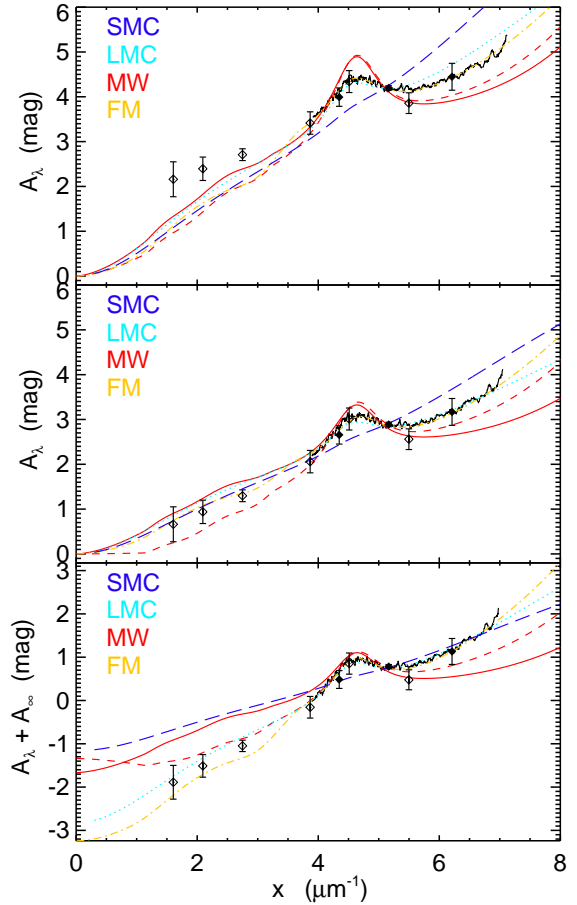}
\caption{The derived extinction curve (black line) of the \grb\ host galaxy.  The curve has been smoothed for clarity in presentation.  The curve has been fit by an SMC (blue long-dashed line), LMC (cyan dotted line), MW with $R_V$ fixed (red solid line) and allowed to vary (red dashed line) and finally using the FM parametrization (yellow dash-dot line).  The filled diamonds are the photometric points presented in this paper used to scale the spectrum while the empty diamonds are the photometric points of \citet{kruhler08}.   Note that the fits were done using only the spectroscopic data.  {\it Top panel:} The derived extinction curve assuming no cooling break between the optical and X-rays in the intrinsic spectrum of the burst, i.e. $\beta=\beta_X=1.02$.    The best fit is obtained using the FM parametrization.  Of the three local extinction laws (MW, SMC, LMC), the LMC provides the best fit.  The photometric points in the infrared are in poor agreement with the extrapolations of the fits and would require a break in the extinction curve at $x\lesssim1.5$~$\mu$m$^{-1}$.  {\it Middle panel:}   The same as the top panel, but using the 1~$\sigma$ deviation value for $\beta$ towards a shallower intrinsic slope, $\beta=0.87$.  In this case, the photometric points are in better agreement with the predicted extinction curve and the sharp break in steepness at around $x\lesssim1.5$~$\mu$m$^{-1}$ is not required.  However, it implies $\beta_X<1$.  {\it Bottom panel:}  The derived extinction curve (black line) assuming a cooling break in the intrinsic spectrum, i.e. $\beta=\beta_X-0.5=0.52$, and shifted by a constant $A_{\infty}$ which depends on the location of the cooling break.  The FM parametrization gives the best fit to the data while the LMC provides the best fit amoung the three local extinction laws.  The photometric points of \citet{kruhler08} are inconsistent with the extrapolation of the SMC and MW (with a fixed $R_V=3.1$), but are in agreement with the other fits.}
\label{fig:extnobreak}
\end{figure}

Fig.~\ref{fig:xraynobreak}
shows the X-ray spectrum (scaled to the SED time, $t_0+3517$) and its
extrapolation into the optical.  The fluxed optical spectrum (mean
time $t_0+2.2$\,hours) was scaled to the $R$-band data point. It is
worth noting that the scaled optical spectrum agrees with the observed
$V$ and $I$ photometric data, indicating that the small temporal
extrapolations are correct and that there is no significant spectral
change in the optical in this period. The extinction curve, shown in
the top panel of Fig.~\ref{fig:extnobreak}, is then derived from
the difference between the spectrum and the extrapolated slope.

To study the properties of the extinction curve, we fit it to the four previously mentioned extinction laws 
with the additional constraint that the extinction must go to zero for infinite wavelengths.  The derived extinction and the fits are shown in the top panel of Fig.~\ref{fig:extnobreak} and the parameters of the fit are given in Tables~\ref{tab:extfits} and \ref{tab:extfitsfm}.  The freedom of the FM parametrization allows for a good fit ($\chi^2/\mathrm{d.o.f.}=261/968$), the LMC also provides a reasonable fit ($\chi^2/\mathrm{d.o.f.}=815/975$), while the SMC and MW fail to reproduce the extinction curve.  However, the infrared photometry points of \citet{kruhler08} clearly do not agree with any of the fits and require the extinction curve to take a very sharp break in its steepness at $x<1.5$~$\mu$m$^{-1}$ which would suggest that the host of \grb\, has a very strange form of extinction law.  However, a more plausible explanation is that the assumed intrinsic slope is wrong, either due to the extrapolation or because of the presence of a cooling break in the spectrum (see \S~\ref{sec:extbreak}).
\begin{table}
\caption{Parameters of the fits with no cooling break}
\begin{tabular}{@{}cccccr@{/}l@{}}
\hline\hline
$\beta$ & Type   & $A_V$   & $R_V$ & $A_V^{\mathrm{fore}}$ & $\chi^2$ &dof\\
\hline
{\it 1.02}& MW &0.87$\pm$0.03 & 1.8$\pm$0.1 & \nodata & $        2281$ & $         974$ \\
{\it 1.02}& MW &1.533$\pm$0.002 & {\it 3.1} & \nodata & $        2755$ & $         975$ \\
{\it 1.02}& FM &1.341$\pm$0.002 & {\it 3.1} & \nodata & $         261$ & $         968$ \\
{\it 1.02}& LMC &1.474$\pm$0.002 & \nodata & \nodata & $         815$ & $        975$ \\
{\it 1.02}& SMC &1.269$\pm$0.002 & \nodata & \nodata  & $        7848$ & $        975$ \\
{\it 1.02} & MW+SMC & $1.274\pm0.002$  & {\it 3.1} & $(0.25) $ & $ 1102$ & $       974$\\
{\it 1.02} & LMC+SMC & $1.474\pm0.002$ & \nodata & $(0) $ & $        815$& $         974$\\[3pt]
{\it 1.19}& MW &1.92$\pm$0.03 & 2.9$\pm$0.1 & \nodata & $        4670$ &         $ 974$ \\
{\it 1.19}& MW &2.052$\pm$0.002 & {\it 3.1} & \nodata & $        4690$ & $         975$ \\
{\it 1.19}& FM &1.820$\pm$0.002 & {\it 3.1} & \nodata & $         321$ & $         968$ \\
{\it 1.19}& LMC &1.976$\pm$0.002 & \nodata & \nodata & $        2654$ & $         975$ \\
{\it 1.19}& SMC &1.705$\pm$0.002 & \nodata & \nodata  & $       16698$ & $        975$ \\
{\it 1.19} & MW+SMC & $1.795\pm0.002$  & {\it 3.1} & $(0.25) $ & $        2821$& $        974$\\
{\it 1.19} & LMC+SMC & $1.976\pm0.002$ & \nodata & $(0) $ & $        2654$& $        974$\\[3pt]
{\it 0.87}& MW &0.32$\pm$0.03 & 0.9$\pm$0.1 & \nodata & $        1229$ & $         974$ \\
{\it 0.87}& MW &1.052$\pm$0.002 & {\it 3.1} & \nodata & $        1859$ & $         975$ \\
{\it 0.87}& FM &0.905$\pm$0.002 & {\it 3.1} & \nodata & $         287$ & $         968$ \\
{\it 0.87}& LMC &1.018$\pm$0.002 & \nodata & \nodata & $         338$ & $         975$ \\
{\it 0.87}& SMC &0.883$\pm$0.002 & \nodata & \nodata  & $        2692$ & $         975$ \\
{\it 0.87} & MW+SMC & $0.796\pm0.002$  & {\it 3.1} & $(0.25) $ & $         649$&$         974$\\
{\it 0.87} & LMC+SMC & $1.018\pm0.002$ & \nodata & $(0) $ & $        338$& $        974$\\
\hline
\end{tabular}
\tablecomments{The parameters of the fits for the different extinction laws assuming no
cooling break between the optical and X-rays in the spectrum.  The slope in
the optical is then $\beta=\beta_X=1.02^{+0.17}_{-0.15}.$ Numbers in italics
were kept fixed, while numbers in brackets reached the limits of their
allowed range.  The error bars are the formal 1~$\sigma$ errors from the
$\chi^2$ minimization.  For the fits assuming a foreground contribution (at
$z=2.08$), $A_V$ is the extinction of the host and $A_V^{\mathrm{fore}}$ is
the extinction of the foreground object, both in their respective
restframes.}
\label{tab:extfits} 
\end{table} 

\begin{table*}
\caption{Additional parameters of the FM fits\label{tab:extfitsfm} }
\begin{center}
\begin{tabular}{@{}cccccccc@{}}
\hline\hline
$\beta$ & $c_1$   & $c_2$ &  $c_3$  & $c_4$ & $c_5$ & $x_c$ & $\gamma$\\
\hline
{\it 1.02} &$0.080\pm0.001$& $1.025\pm0.001$& $2.747\pm0.007$& $0.355\pm0.003$ & $5.239\pm0.013$& $4.583\pm0.004$& $1.084\pm0.002$ \\
{\it 1.19} &$0.072\pm0.001$& $1.033\pm0.001$& $2.719\pm0.005$& $0.278\pm0.005$ & $5.889\pm0.031$& $4.463\pm0.003$& $1.134\pm0.002$ \\
{\it 0.87} &$0.079\pm0.001$& $1.011\pm0.002$& $2.823\pm0.010$& $0.531\pm0.005$ & $4.779\pm0.012$& $4.617\pm0.007$& $1.040\pm0.003$ \\[6pt]

{\it 0.52} &$0.083\pm0.001$& $1.018\pm0.002$& $2.683\pm0.007$& $0.317\pm0.002$ & $4.672\pm0.010$& $4.642\pm0.005$& $1.094\pm0.002$ \\
{\it 0.69} &$0.071\pm0.003$& $1.027\pm0.001$& $2.723\pm0.008$& $0.366\pm0.002$ & $4.847\pm0.012$& $4.606\pm0.005$& $1.095\pm0.002$ \\
{\it 0.37} &$0.073\pm0.001$& $1.023\pm0.002$& $2.837\pm0.012$& $0.440\pm0.003$ & $4.160\pm0.011$& $4.667\pm0.008$& $1.065\pm0.004$ \\
\hline
\end{tabular}
\end{center}
\tablecomments{The additional parameters of the FM parametrization for fits
with or without a cooling break. The errors quoted are the formal $1\sigma$
errors.  The uncertainty in the parameters is dominated by the uncertainty
in $\beta$.  The top three lines correspond the $\beta=\beta_X=1.02$ and the
$1~\sigma$ deviations, while the lower three lines correspond to
$\beta=\beta_X-0.5=0.52$ and the $1~\sigma$ deviations. }
\end{table*}

From Fig.~\ref{fig:xraynobreak} we see that due to the several order of magnitudes extrapolation, the 1~$\sigma$ difference in $\beta$ translates into a difference of $\sim$1 magnitude in the optical.  While going towards steeper slopes will make the apparent 'break' in the extinction curve greater, a shallower slope will act to reduce it as seen in Fig.~\ref{fig:extnobreak}.  Redoing the analysis for a shallower slope corresponding to 1~$\sigma$ deviation (i.e. $\beta=0.87$) results in the extinction curve shown in the middle panel of Fig.~\ref{fig:extnobreak}.    The parameters of the fit, for both 1~$\sigma$ deviations are given in Tables~\ref{tab:extfits} and \ref{tab:extfitsfm}.  The goodness of fit for the MW, SMC and LMC is improved for the shallower slope, while the effect on the FM fit is minimal.  The fits  are also more consistent with the additional photometric points in the infrared of \citet{kruhler08}.  We note however that most GRBs have $\beta_X > 1$ \citep[see e.g.,][ their fig.~3, with $\Gamma=\beta+1$]{2007ApJ...670..565L}.

\subsection{The extinction curve assuming a cooling break}
\label{sec:extbreak}
A cooling break in the intrinsic spectrum occurring between the X-ray and the optical data, i.e. $\beta=\beta_X-0.5$, would result in an intrinsic spectrum at optical wavelengths of the form
\begin{equation}
f^{\mathrm{intr}}_{\nu}=\tilde{f}_{\nu,0}\left(\frac{x}{x_0}\right)^{-\beta}=\tilde{f}_{\nu,0}\left(\frac{x}{x_0}\right)^{-(\beta_X-0.5)}
\end{equation}
where the normalization $\tilde{f}_{\nu,0}$ depends on the location of the cooling break, $x_{\mathrm{break}}$, by
\begin{equation}
x_{\mathrm{break}}=x_0\left(\frac{f_{\nu,0}}{\tilde{f}_{\nu,0}}\right)^{2}.
\end{equation}  
The absolute extinction is then given by
\begin{eqnarray}
A_{\lambda}&=&-2.5\log_{10}\left(\frac{f_{\nu}}{\tilde{f}_{\nu,0}}\left(\frac{x}{x_{0}}\right)^{\beta}\right)\\
&=&-2.5\log_{10}\left(\frac{f_\nu}{f_{\nu,0}}\left(\frac{x}{x_0}\right)^{\beta}\right) - A_{\infty}.
\end{eqnarray}
The first term of the sum can be determined as before from the X-ray data, while the second part of the sum 
\begin{equation}
A_{\infty}\equiv2.5\log_{10}\frac{f_{\nu,0}}{\tilde{f}_{\nu,0}}=2.5\log_{10}\left(\frac{x_{\mathrm{break}}}{x_0}\right)^{0.5}
\end{equation}
is an additional unknown to be added to the fits.  Its effect is to shift the extinction curve up or down (i.e., the value of $A_\lambda + A_\infty$ at $x=0$ is $A_\infty$).  Constraining the break to occur between the optical and X-ray regimes translates into constraints on the allowed values of $A_\infty$ corresponding to setting $x_{\mathrm{break}}$ to be the upper $x$ of the optical data and the lower $x$ of the X-ray data.  An additional requirement is that the extinction remains positive.

We fit the three local extinction laws (MW, SMC and LMC) to the curve in addition to the FM parametrization.  Again we find that of the three local extinction laws the LMC provides the best fit ($\chi^2/\mathrm{d.o.f.}=311/974$), but the FM parametrization still gives the best overall fit ($\chi^2/\mathrm{d.o.f.}=253/967$).   In this instance, the infrared photometric points of \citet{kruhler08} agree with the extrapolated LMC and FM fits.

\subsection{Possible contribution from a foreground object}
\label{sec:ext_fore}
\begin{figure}
\includegraphics[angle=0,width=\columnwidth,clip=]{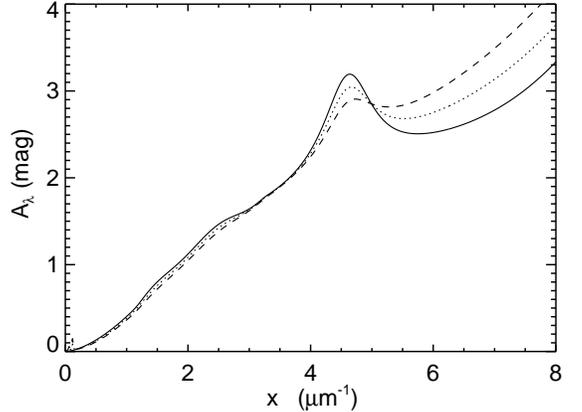}
\caption{The effect on an "observed" extinction curve assuming that the extinction is caused by a MW type extinction in addition to an SMC foreground object (located at $z=2.45$ and $z=2.08$ respectively).  The three lines correspond to $A_V=1.0$, $A_V^{\mathrm{fore}}=0.0$ (solid line), $A_V=0.75$, $A_V^{\mathrm{fore}}=0.25$ (dotted line) and $A_V=0.5$, $A_V^{\mathrm{fore}}=0.5$ (dashed line) where the $A_V$ values are quoted in their restframes.}
\label{fig:ext_effect_foreground}
\end{figure}

The MW extinction law provides a poor fit to the observed extinction, in part because the observed bump is shallower than the predicted MW bump and the rise into the UV is steeper (see e.g. Fig.~\ref{fig:extnobreak}).  However, as Fig.~\ref{fig:ext_effect_foreground} shows, such effects can be caused if part of the extinction is due to a foreground object without a bump.

The spectrum of \grb~shows two intervening \ion{Mg}{2} absorption systems which might also cause part of the reddening.  Neither foreground system shows any signs of a $2175$~\AA\ bump.  If the foreground absorber had a $2175$~\AA\ bump in its restframe, it would act to broaden and shift the peak of the observed bump in the restframe of the host, which is inconsistent with the data.  Therefore, the most straightforward way to model them is with an SMC type extinction law.  As this extinction law is roughly linear, there is no way of separating the effect of two systems without additional constraints on the dust content of the two systems.   As the absorber at $z=2.29$ has much weaker absorption lines, and therefore by assumption much weaker extinction, we will only look at the possible contribution from the system at $z=2.078$.   

We will therefore limit ourselves to studying the possibility of the extinction being of the form 
\begin{equation}
A_\lambda=A_\lambda^{\mathrm{host}}+A_\lambda^{\mathrm{fore}}
\end{equation}
where $A_\lambda^{\mathrm{host}}$ is the extinction in the host galaxy and $A_\lambda^{\mathrm{fore}}$ is the extinction in the foreground \ion{Mg}{2} absorber.   We have already seen that the extinction curve is poorly fit by an SMC type extinction, as it lacks the characteristic bump.  We will therefore parametrize the extinction of the host galaxy of \grb\, as either a MW (with fixed $R_V=3.1$) or an LMC type, while the foreground absorber will be parametrized as an SMC type.   In addition, we place the limit $A_V^{\mathrm{fore}}<0.25$ for the foreground absorber based on the discussion in \S~\ref{sec:foreground}.

\begin{figure}
\includegraphics[angle=0,width=\columnwidth,clip=]{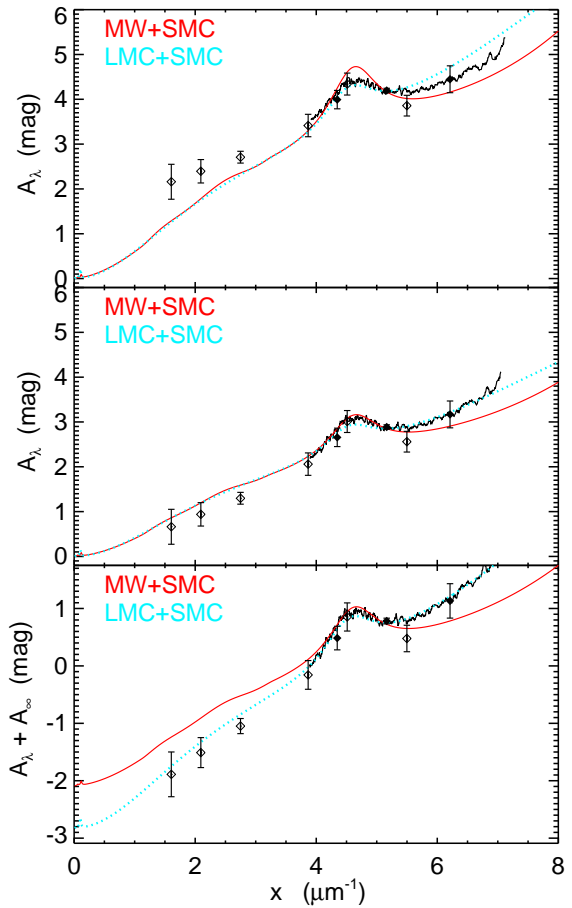}
\caption{Extinction curve fits taking into account a possible contribution from the foreground absorber at $z=2.078$.  The extinction of the host is taken to be a MW (red solid line) or an LMC (cyan dotted line).   The extinction of the foreground absorber is taken to be an SMC (as there is no evidence of a bump in the data) with an upper limit of $A_V^{\mathrm{fore}}\le0.25$ (see \S~\ref{sec:foreground}).  The filled diamonds are the photometric points presented in this paper used to scale the spectrum while the empty diamonds are the photometric points of \citet{kruhler08}.  Note that the fits were done using only the spectroscopic data.   The top panel shows the derived extinction assuming no cooling break.  The middle panel shows the 1~$\sigma$ deviation towards a shallower intrinsic slope.  The bottom panel shows the derived extinction assuming a cooling break.  For the LMC fits the contribution of the foreground extinction is not found to improve the fits (setting $A_V^{\mathrm{fore}}=0$).  The MW fits are improved by adding an SMC contribution, however, they are still worse than LMC only, and would require $A_V^{\mathrm{fore}}>0.25$.}
\label{fig:extnobreak_double}
\end{figure}

The results are shown in Fig.~\ref{fig:extnobreak_double} and tabulated in Tables~\ref{tab:extfits} and \ref{tab:extfitsbreak}.  We find that for the LMC, the addition of the foreground contributor does not improve the fits and the best fits are found by setting $A_V^{\mathrm{fore}}$ to be zero.  For the MW, the fits are improved and the best fit is found by setting $A_V^{\mathrm{fore}}$ to its maximum value of $0.25$.  The MW fits are still worse than the LMC leading us to conclude that the extinction of \grb~is not well fitted by a MW type extinction. 

\begin{table*}
\caption{Parameters of the fits assuming a cooling break\label{tab:extfitsbreak}}
\begin{center}
\begin{tabular}{@{}llllllr@{/}l@{}}
\hline\hline
\multicolumn{1}{c}{$\beta$} & Type   & $A_V$ &  $A_\infty$  & $R_V$& $A_V^{\mathrm{fore}}$&$\chi^2$& dof\\
\hline
{\it 0.52}& MW & (0) & $-1.47\pm$0.04 & 0$\pm$0.003 & \nodata & $   938$ & $   973$ \\
{\it 0.52}& MW &0.68$\pm$0.02 & $-1.14\pm$0.06 & {\it 3.1}& \nodata & $  2517$ & $   974$ \\
{\it 0.52}& FM &1.259$\pm$0.002 & ($-3.2$) & {\it 3.1}& \nodata & $  253$ & $   967$ \\
{\it 0.52}& LMC &1.27$\pm$0.02 & $-2.86\pm$0.07 & \nodata & \nodata & $   311$ & $   974$ \\
{\it 0.52}& SMC &0.52$\pm$0.01 & $-0.97\pm$0.04 & \nodata & \nodata  & $  1078$ & $   974$ \\
{\it 0.52} & MW+SMC & $0.69\pm0.02 $ & $-1.85\pm0.06 $ & {\it 3.1}& (0.25) &   1030&   973\\
{\it 0.52} & LMC+SMC & $1.27\pm0.02 $ & $-2.86\pm0.07 $ & \nodata & (0) &    311& 973\\[3pt]
{\it 0.69}& MW & (0) & ($-0.5$) & 0$\pm$0.01 & \nodata & $   994$ & $   973$ \\
{\it 0.69}& MW &0.964$\pm$0.002 & ($-0.5$) & {\it 3.1}& \nodata & $  2155$ & $   974$ \\
{\it 0.69}& FM &1.25$\pm$0.03 & $-1.76\pm$0.08 & {\it 3.1}& \nodata & $253$ & $   967$ \\
{\it 0.69}& LMC &1.17$\pm$0.02 & $-1.17\pm$0.06 & \nodata & \nodata & $ 307$ & $ 974$\\
{\it 0.69}& SMC &0.807$\pm$0.002 & ($-0.5$) & \nodata & \nodata  & $  1956$ & $   974$ \\
{\it 0.69} & MW+SMC & $0.706\pm0.002 $ & ($-0.5$) & {\it 3.1}& (0.25)  &    784&   973\\
{\it 0.69} & LMC+SMC & $1.17\pm0.02 $ & $-1.17\pm0.06 $ & \nodata & (0) &    307& 973\\[3pt]
{\it 0.37}& MW &(0) & $-2.67\pm$0.04 & $0\pm0.005$& \nodata & $  1125$ & $   973$ \\
{\it 0.37}& MW &0.73$\pm$0.02 & $-2.57\pm$0.06 & {\it 3.1}& \nodata & $  2671$ & $   974$ \\
{\it 0.37}& FM &0.807$\pm$0.002 & ($-3.2$) & {\it 3.1}& \nodata & $   410$ & $   967$ \\
{\it 0.37}& LMC &0.937$\pm$0.002 & ($-3.2$) & \nodata & \nodata & $   559$ & $   974$ \\
{\it 0.37}& SMC &0.58$\pm$0.01 & $-2.45\pm$0.04 & \nodata & \nodata  & $  1060$ & $   974$ \\
{\it 0.37} & MW+SMC & $0.69\pm0.02 $ & $-3.15\pm0.06 $ & {\it 3.1}&(0.25) &   1184&   973\\
{\it 0.37} & LMC+SMC & $0.78\pm0.02 $ & ($-3.2$) & \nodata & $0.16\pm0.02 $ & 512&  973\\
\hline
\end{tabular}
\end{center}
\tablecomments{The parameters of the fits for the different extinction laws assuming a cooling break between the optical and the X-rays in the spectrum.  The slope in the optical is then $\beta=\beta_X-0.5=0.52^{+0.17}_{-0.15}.$  Numbers in italics were kept fixed, while numbers in brackets reached the limits of their allowed range.  The error bars are the formal 1~$\sigma$ errors from the $\chi^2$ minimization.   For the fits assuming a foreground contribution (at $z=2.08$), $A_V$ is the extinction of the host and $A_V^{\mathrm{fore}}$ is the extinction of the foreground object, both in their respective restframes.}
\end{table*}

\subsection{Summary of the extinction curve properties}
\begin{figure*}
\begin{center}
\includegraphics[angle=0,width=0.6\textwidth,clip=]{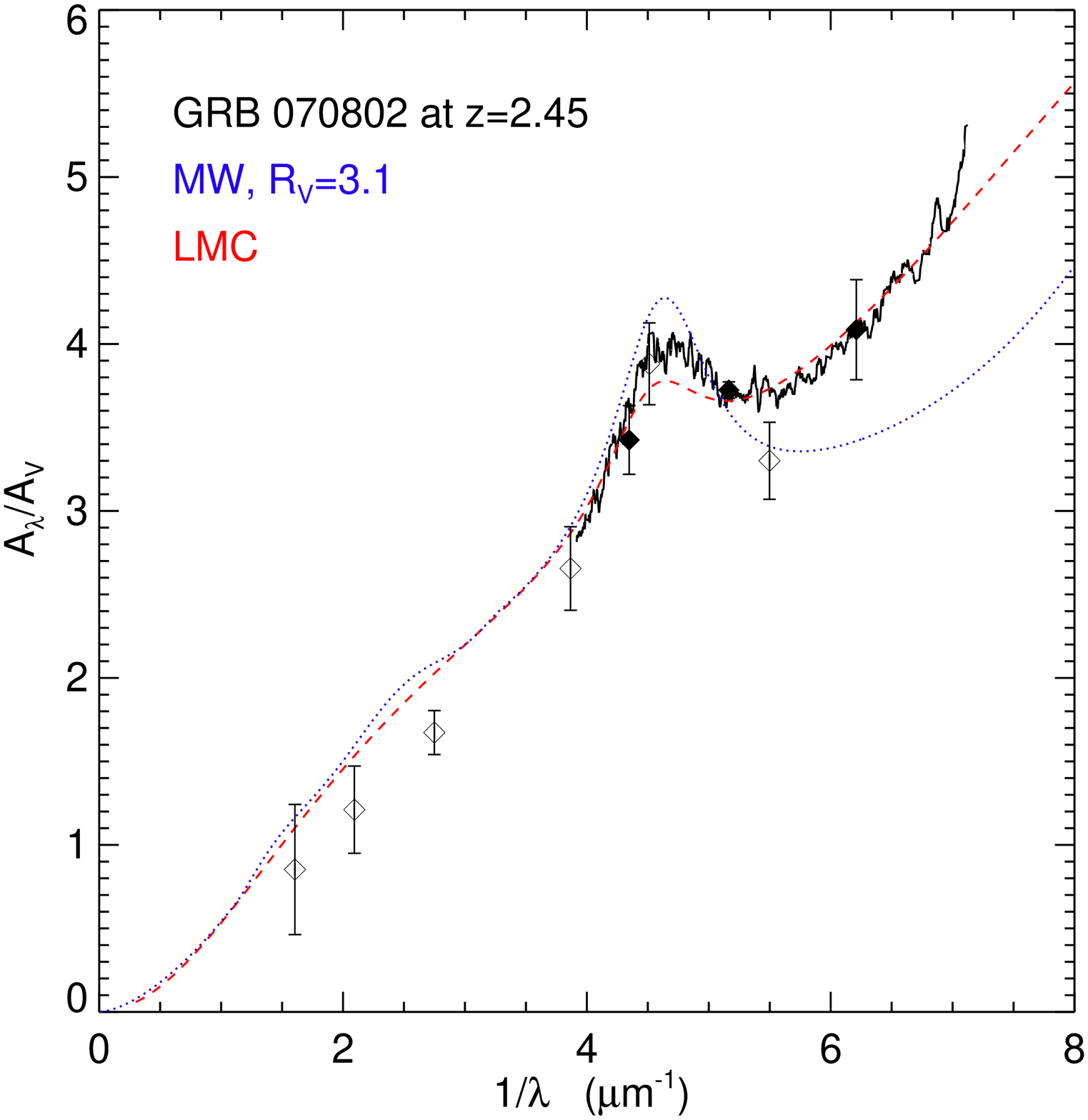}
\end{center}
\caption{An absolute extinction curve for the afterglow \grb\ at $z=2.45$.  It is based on the simplest model of the afterglow that is consistent with all the available data (single power-law spectrum, $\beta = 0.87$, corresponding to a $1$~$\sigma$ deviation from the best fit) and with the constraint that the extinction should be zero at $1/\lambda=0$.  The extinction curve shows a clear bump at $2175$~\AA.   Also shown are the best fits (to the spectroscopic data) for an LMC extinction (dashed red line) and MW extinction (dotted blue line), with the LMC clearly providing the better fit. To obtain the absolute extinction, A$_V$ was derived using a linear interpolation of the two datapoints closest to the restframe $V$-band.  The filled diamonds are the photometric points presented in this paper used to scale the spectrum while the empty diamonds are the photometric points of \citet{kruhler08}. }
\label{fig:ext_clear}
\end{figure*}

Fig.~\ref{fig:ext_clear} shows the scaled extinction curve for \grb\ for $\beta=0.87$.
We clearly detect the 2175~\AA\ bump in the extinction curve of \grb.  The data sample the bump very well and its detection does not depend on whether we assume cooling breaks or not in the spectrum of the GRB.  It is one of the most robust detections of the bump in extragalactic environments to date and currently the highest redshift detection at z=2.45 (corresponding to 2.5 Gyr assuming a flat universe with $\Omega_m=0.3$, $\Omega_\Lambda=0.7$ and $H_0=73$~km~s$^{-1}$~Mpc$^{-1}$).

For the three local extinction laws, the LMC provides the best fit (i.e., it has the lowest $\chi^2$ per degree of freedom (d.o.f.)) to the shape of the derived extinction curve, regardless of whether we assume there is a cooling break or not between the X-rays and the optical, and whether we take into account a possible contribution from the foreground absorbers or not (see Tables~\ref{tab:extfits} and \ref{tab:extfitsbreak}).  The SMC clearly provides a poor fit in all cases as the bump is completely missing, while the bump of the MW extinction law is too strong and the rise into the UV not steep enough.  A foreground absorber with SMC type extinction could make the bump shallower and the UV rise steeper (see Fig.~\ref{fig:ext_effect_foreground}), however, the absorption would need to be larger than the upper limits placed on $A_V^{\mathrm{fore}}$ in \S~\ref{sec:foreground}.

The best fits are obtained using the FM parametrization.  This is not surprising, as it has more freedom in tracing the shape of the curve (see Appendix~\ref{app:ext}).  The three parameters giving the shape of the bump are only weakly dependent on the assumed $\beta$ and whether we assume a cooling break or not (Table~\ref{tab:extfitsfm}).  The bump is found to be centered at $x_c\approx4.6\pm0.1$~$\mu$m$^{-1}$, the width of the bump is found to be $\gamma\approx1.08\pm0.05$~$\mu$m$^{-1}$ while the 'strength' (i.e. its height above the linear extinction, see Appendix~\ref{app:ext}) is $c_3\approx2.7\pm0.1$.  Therefore, both the area of the bump $\Delta_\mathrm{bump}\equiv \pi c_3/(2\gamma)$ and its maximum height above the linear extinction $E_{\mathrm{bump}}\equiv c_3/\gamma^2$ are well defined.  The relative strength of the bump $A_{\mathrm{bump}}/A_V=c_3/(\gamma^2 R_V)$ \citep[as defined by ][]{gordon2003} is also well defined, although here the uncertainty is dominated by the uncertainty in $A_V$ (or equivalently $R_V$).  We note that the value of $c_3$ is the same as the average $c_3=2.7\pm0.1$ that \citet{gordon2003} find for the LMC average sample while the width $\gamma$ is a bit wider compared to their $\gamma=0.93\pm0.02$~$\mu$m$^{-1}$ (although it is still within their scatter).  The value of $c_3$ is however higher than \citet{gordon2003} find for the LMC2 sample ($c_3=1.5\pm0.1$) although it is within the scatter.

We estimate the amount of dust extinction to be $A_V=1.341\pm0.002$ for $\beta=1.02$ and $A_V=1.259\pm0.002$ for $\beta=0.52$ given the FM parametrization.  The LMC fits correspondingly give $A_V=1.474\pm0.002$ for $\beta=1.02$ and $A_V=1.27\pm0.02$ for $\beta=0.52$.  Taking into account the possible $1~\sigma$ deviation of the slope \citep[excluding steeper slopes than $\beta=1.02$, as this would be in strong disagreement with the photometric points of][]{kruhler08}, we estimate $A_V=0.8$--$1.5$ in the host along the line of sight to \grb.

\section{Discussion}
\label{sec:dis}
The nature of the interstellar extinction peak at 2175\,{\AA} remains
poorly understood more than 40 years after its discovery by \citet{stecher}.  
As its detection has been limited to the Milky Way with only a few exceptions 
it has proven hard to search for correlations between the dust environment 
and the detection or non-detection of the bump.  The detection of the bump 
in the spectrum of \grb~is interesting in itself for two reasons, i.e., as 
being the highest redshift detection of the bump and as being the first 
robust detection of the bump in a GRB host galaxy. It shows that the carrier of the 2175\,{\AA} bump, which is characteristic for Milky Way
type dust, was in place 2.6 Gyr after the Big Bang (when the Universe was
only 20~\% of its current age).  It also shows that the conditions for both forming and
not destroying the 2175\,{\AA} were satisfied in a GRB host galaxy -- surprising
in view of the fact that most GRB host galaxies are faint, blue, young,
low-metallicity galaxies
\citep{2003A&A...400..499L,2004A&A...425..913C,2006Natur.441..463F} in contrast to the massive, evolved and 
chemically enriched Milky Way.

Moreover, the detection of 2175\,{\AA} as well as a very rich spectrum
of redshifted UV metal absorption lines allows us to explore various 
hypotheses for the origin of the bump -- an experiment that cannot 
easily be done along lines of sight in the Milky Way. Below we explore
whether there are any other properties of the galaxy which are correlated 
with the presence of the bump.  We discuss what separates \grb~from other 
GRBs with reddening but no detected bump and compare the properties of the 
host galaxy with other galaxies for which the extinction curve has been 
determined.

\subsection{Metallicity}
As mentioned in \S~\ref{sec:2175},
based on evidence from the MW, LMC and SMC, the simplest hypothesis would be
that the strength of the 2175 \AA\  bump is simply controlled by the
metallicity.  However, 
\citet{savaglio2003}  infer a metallicity of 
[Zn/H]$ = -0.13\pm0.25$ for GRB\,000926  (compared to [Zn/H]$=-0.50\pm0.68$ we derive for \grb) and the \ion{H}{1} column density for GRB\,000926  is 
similar to that of GRB\,070802.  This would indicate that the extinction curve of GRB\,000926 should also have a bump if metallicity was its only driver.  However, the extinction curve derived for 
GRB\,000926 is inconsistent with those of the MW and LMC and fully consistent 
with that of the SMC, i.e., without a 2175 \AA\  bump \citep{2001A&A...373..796F}.

\subsection{Gas to dust ratio}
\citet{gordon2003} suggested that the average extinction properties of the SMC, LMC and MW were described by the same underlying extinction law, which could be described by the $R_V$ parameter of the MW extinction and one additional parameter characterizing the strength of the bump and the UV slope.  They suggested the gas to dust ratio as the second parameter and showed that for the LMC and SMC there is a correlation between the gas to dust ratio and the UV slope (i.e. $c_2/R_V$) and an anti-correlation between the gas to dust ratio and the strength of the bump (i.e. $ A_{\mathrm{bump}}/A_V=c_3/(\gamma^2 R_V)$), albeit with a large scatter.

\begin{figure*}
\begin{center}
\includegraphics[angle=0,width=0.8\textwidth,clip=]{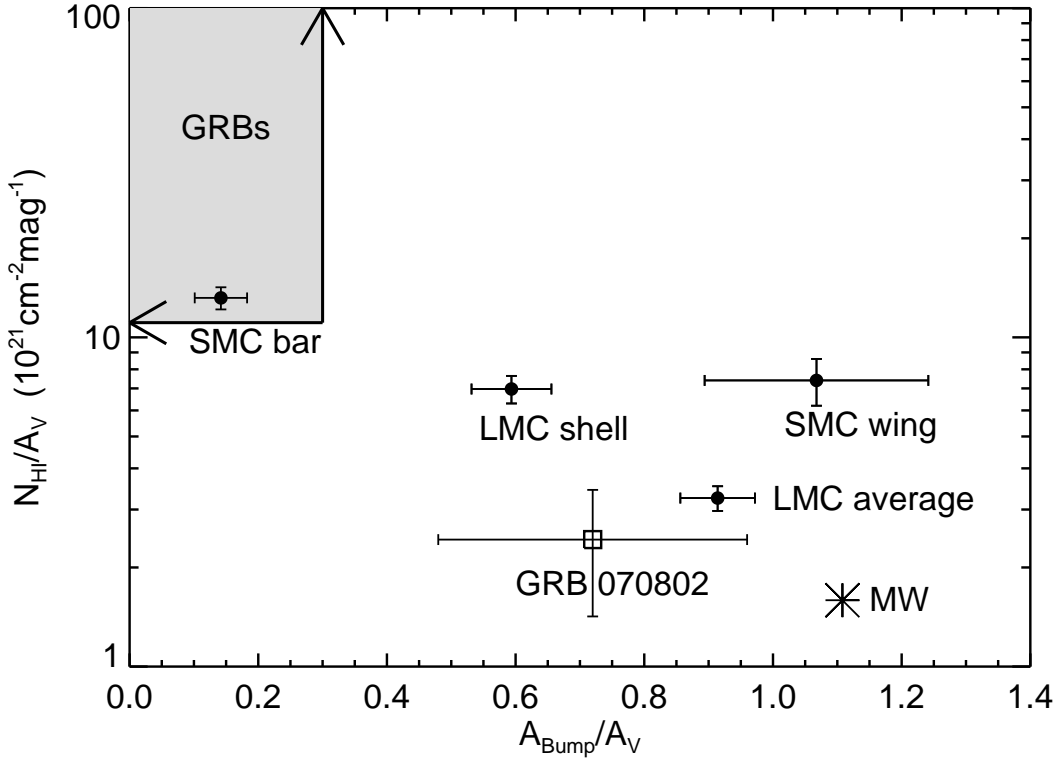}
\end{center}
\caption{ $N_H/A_V$ vs. $A_{\mathrm{bump}}/A_V$.  The black filled circles are the averaged values for the SMC (bar and wing (only one line of sight)) and LMC (shell and average) from \citet{gordon2003}.  The MW is marked with a star.  \grb~is marked by a square, while other GRBs from the literature fall into the shaded region bounded by the arrows in the upper left corner.  The $N_H/A_V$ values for the GRBs are calculated from the values quoted in Table~\ref{tab:n_H}.  Their $A_{\mathrm{bump}}/A_V$ values are estimated to be lower than $0.3$ as they are all consistent with SMC type extinction (i.e. no sign of a bump in the SED).   The values for \grb~are calculated directly from the parameters of the \citet{fitzpatrick2007} parametrization (see Table~\ref{tab:extfitsfm}).   The error bars are dominated by the uncertainty in $A_V$ (and equivalently $R_V$), for which we have taken $A_V=1.3\pm0.5$ (or $R_V=3.1\pm1.2$).  The $A_{\mathrm{bump}}/A_V$ for the MW is obtained by fitting an FM curve to a standard MW extinction (i.e. $R_V=3.1$) and $N_H/A_V=4.93/R_V\cdot10^{21}$~cm$^{-2}$~mag$^{-1}$ is taken from \citet{1994ApJ...427..274D}.  We find that the trend suggested by \citet{gordon2003} of decreasing bump strength with increasing gas-to-dust ratio holds for the MW, \grb\ and the other GRBs from the literature.}
\label{fig:n_H}
\end{figure*}

GRBs typically originate in host galaxies with low dust to gas ratios \citep[i.e. high gas to dust ratios][]{2003ApJ...597..699H,jakobsson2004,2004A&A...419..927V,2006ApJ...641..993K,2007ApJ...666..267P,2007MNRAS.377..273S}.  While the column density, $N_H$, of the host of GRB\,070802 is not exceptionally large, it does have a very large $A_V$, leading to a low gas to dust.  This is shown in Fig.~\ref{fig:n_H}, where we have plotted $N_H/A_V$ vs. the strength of the bump $A_{\mathrm{bump}}/A_V$ for the SMC, LMC, MW and \grb.  Also plotted is the region where GRBs for which an extinction curve analysis exists in the literature lie.  The plot shows that the anti-correlation between the bump strength and the gas to dust ratio found by \citet{gordon2003} for the SMC and the LMC also holds for the GRBs, and \grb\ in particular, although with large uncertainties in the values.
\begin{table}
\caption{Extinction and column density of GRBs\label{tab:n_H}}
\begin{center}
\begin{tabular}{@{}lllll@{}}
\hline\hline
\multicolumn{1}{c}{Name}   & \multicolumn{1}{c}{$A_V$} &  \multicolumn{1}{c}{$\log_{10} N$(\ion{H}{1})} & \multicolumn{1}{c}{$z$}&\multicolumn{1}{c}{References}\\
\hline
GRB~000301C &  $0.09\pm0.04$		& $21.2\pm0.5$	& 2.040 &	 1	\\
GRB~000926	&  $0.18\pm0.06$		& $21.3\pm0.2$	& 2.038 & 2,3	\\
GRB~020124 	&  $<0.2$			& $21.7\pm0.2$	& 3.198 &	 4 	\\
GRB~030323	&  $<0.5$				& $21.90\pm0.07$	& 3.372 &	 5	\\
GRB~030429 	&  $0.34\pm0.04$		& $21.6\pm0.2$	& 2.658 &	 6	\\
GRB~050401	&  $0.62\pm0.06$		& $22.6\pm0.3$	& 2.899 &	 7	\\
\grb			& $0.8-1.5$			&$21.5\pm0.2$		& 2.455 & 	8	\\
\hline 
\end{tabular}
\end{center}
\tablecomments{The GRBs represented in Figure~\ref{fig:n_H} as the shaded region and \grb.  This is not a homogenous sample with differing datasets and analysis from the literature.}
\tablerefs{ (1) \citet{2001A&A...370..909J} (2) \citet{2002luml.conf..187F} (3) \citet{2001A&A...373..796F}  (4) \citet{2003ApJ...597..699H} (5) \citet{2004A&A...419..927V}  (6) \citet{2004A&A...427..785J}  (7)\citet{2006ApJ...652.1011W}  (8) This paper. }
\end{table}

\subsection{Detection of \ion{C}{1} and the UV radiation field}
\label{sec:CI}
\grb\ is to our knowledge the first GRB to date to show 
prominent \ion{C}{1} in its afterglow spectrum and also the first GRB to show the
2175\,{\AA} feature, which is suggestive of a link between these two components. 
Should such a link be established by
further observations it constitutes a constraint on the environment in which the
2175\,{\AA} feature occurs. The ionization potential of
neutral carbon is lower than that of hydrogen
(11.3 vs.\ 13.6 eV) and neutral carbon is therefore only expected in regions
without intense UV radiation.  This would be consistent with the general lack
of the 2175\,{\AA} feature in GRB host galaxies, as GRBs arise predominantly
from star-forming regions where the UV radiation is expected to be strong
\citep{2006Natur.441..463F, 2008arXiv0809.2608C}. 

\begin{figure}
\includegraphics[angle=0,width=\columnwidth,clip=]{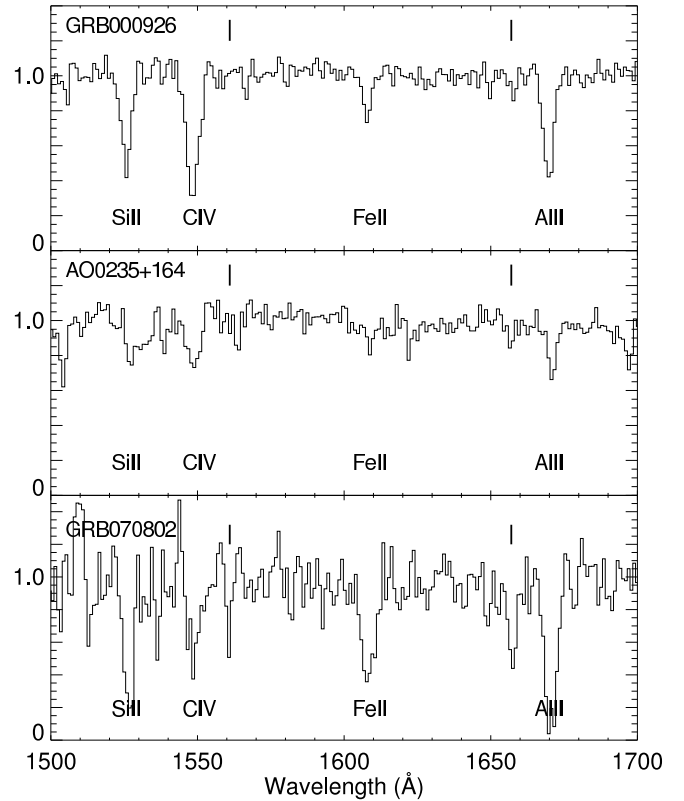}
\caption{A comparison of the region around \ion{C}{1} in the normalized spectra
of GRB~000926, GRB~070802, and AO~0235+164. The vertical lines show the 
positions of the \ion{C}{1} lines.
The two GRB systems have nearly
identical \ion{Si}{2} line strengths, but GRB~000926 has stronger \ion{C}{4}
and no significantly detected  \ion{C}{1}.  \grb\ also has much stronger \ion{Al}{2} and \ion{Fe}{2}
than GRB~000926. AO~0235+164, which has a similar \ion{H}{1} column density as
the two GRB sightlines, displays both the 2175~{\AA} bump and significant \ion{C}{1}.}
\label{fig:26vs02}
\end{figure}

To test whether this suggested correlation holds in general, we have looked at
the other three robust detections of the 2175\,{\AA} bump in individual
systems beyond the local group.  For the lensing galaxy reported by
\citet{motta2002} and the intervening absorber towards GRB~060418 reported by \citet{2006MNRAS.372L..38E}, the existing data do not cover the wavelength range where
one would see the \ion{C}{1} absorption line if present.  However, for the
intervening damped Lyman-$\alpha$ system toward AO~0235+164 reported by
\citet{junkkarinen2004} we see tentative evidence for the \ion{C}{1}\,$\lambda$1656
absorption line. The regions around  \ion{C}{1}\,$\lambda$1560,$\lambda$1656 are shown in
Fig.~\ref{fig:26vs02} for GRB\,070802 and AO~0235+164. For comparison we also
show the same region in the afterglow spectrum of the high metallicity burst
GRB\,000926 which shows no significant evidence for the \ion{C}{1} absorption 
features (and no bump in its extinction curve).  

It would be of interest to further check whether the sightlines to the MW, LMC and SMC for which extinction curve analysis exist are consistent with this correlation.  Such a study is beyond the scope of this current work, but we note that a quick and incomplete search of the literature resulted in two more lines of sight consistent with this correlation.  The first is a detection of \ion{C}{1} in the line of sight towards HD 185418 in the MW by \citet{2003ApJ...596..350S}.  This line of sight is included in the sample of \citet{fitzpatrick+massa86} and has the 2175~\AA\ bump in its extinction curve.  The second is a line of sight towards the SMC, which has \ion{C}{1} detected in the MW but only as an upper limit in the SMC \citep{1997ApJ...489..672W}.  Although this sightline does not have an extinction curve analysis, it is consistent with our prediction that most lines of sight in the SMC should not show significant \ion{C}{1} absorption.

The presence of the \ion{C}{1} absorption feature in GRB\,070802 suggests that the UV radiation field is weaker than in typical GRB environments (see \S~\ref{sec:CI}).  \citet{gordon2003} reached the tentative conclusion that the shape of the extinction curve is affected by the UV flux density in the environment of the dust. In particular, a weaker UV flux density is found to correlate with the presence of the bump.   Continuing our comparison to the bump-less GRB\,000926 host, we find evidence that GRB\,070802 does indeed have a weaker UV radiation field:
{\it i)} the GRB\,000926 absorption system has stronger high ionization 
lines and much weaker lines from neutral species (e.g., \ion{C}{1}, 
see Fig.~\ref{fig:26vs02}) than the GRB\,070802 system. 
 {\it ii)} the host galaxy of GRB\,000926 appears to have a stronger 
 UV flux density as illustrated by the very strong Lyman-$\alpha$ emission
 line \citep{fynbo2002}. 
 For GRB\,070802 we can exclude the presence
 of such a strong Lyman-$\alpha$ emission line (Fig.~\ref{fig:spectrum} and Milvang-Jensen
 et al., in preparation).
 This supports the conclusion reached by \citet{gordon2003}.

It is not immediately clear what the physical significance of the possible
correlation of strong \ion{C}{1} and the 2175\,\AA\ bump is. Given that the
bump is generally believed to be carried by carbonaceous material, and that
carbonaceous grain growth and formation requires free, neutral carbon and
molecules \citep{henning+salama}, it would not be surprising to find both observed properties in
the same environments. The simultaneous presence of \ion{C}{1} and the
2175\,\AA\ bump as well as the lower overall ionization state of the gas,
relatively (though not exceptionally) high metallicity, and large
dust-to-gas ratio may be explained in a scenario in which the dust column is
strongly enriched by the presence of asymptotic giant branch (AGB) stars. 

For massive stars to move onto the AGB requires at least 600\,Myr and
typically much longer for a large population \citep{1992A&A...264..105M}.
The star-forming environment at such an age will be intrinsically relatively
benign, with a softer UV field, and one in which a large amount of dust and
molecular and free carbon is produced
\citep{andersen03,2004A&A...422..289G}. Furthermore the interstellar medium
(ISM) is likely to be reasonably metal-rich and dust-rich. These properties
are in contrast to the normal environments of GRB hosts which are typically
metal-poor. However some GRB hosts may be fairly metal-enriched \citep{2006A&A...451L..47F} but
still have hard radiation fields and young stellar populations \citep{2003A&A...400..499L,2004A&A...425..913C,2004ApJ...611..200P} and in particular, low
dust-to-gas ratios \citep{2006A&A...451L..47F,2006A&A...460L..13J,2007ApJ...660L.101W}.
This is consistent with the above scenario since the metal-enrichment timescale could well be much shorter than the
$\gtrsim10^9$\,yr required to have a reasonable number of AGB stars
producing dust \citep{2004MNRAS.351.1379S}. Such a scenario is then also
consistent with the fact that GRB\,070802 is the only GRB host galaxy so far
discovered with a 2175\,\AA\ bump.  It should also be noted that the host
galaxy of \grb\ is fairly luminous and red for a GRB host
\citep{2008arXiv0803.2718S}, suggesting that it is a massive, evolved system, which would be in agreement with the claim of \citet{noll07} that
the presence of the bump requires an evolved population.

\section{Conclusions}
\label{sec:con}
We have presented  VLT observations of the afterglow of \grb. In a low resolution spectroscopy of the optical afterglow we detect a large number of strong metal lines from absorption systems at $z=2.078$, $z=2.292$ and $z=2.4549$. The highest redshift system is remarkable in showing very strong metal lines, e.g. with higher $W_r$ for the \ion{Si}{2} lines at 1526 and 1808 \AA\ than for any other known 
absorption system we are aware of.  We also detect strong absorption from \ion{C}{1} implying that the gas is shielded from strong UV radiation. The spectrum shows a red wing of a  Ly$\alpha$ line from which we derive a \ion{H}{1} column density of $\log(N($\ion{H}{1}$))=21.50\pm$0.20.  Imaging of the field revealed a fairly bright and red host, detected both in $R$ and $K$ bands, suggesting that it is an evolved, massive galaxy.

The spectrum is also remarkable in that the extinction curve of the line of sight towards \grb\ shows a clear signature of the so called Milky Way or $2175$~\AA\ bump.  At a redshift of $z=2.45$, it is by far the highest redshift detection of the $2175$~\AA\ bump to date.  It shows that the conditions for the creation and non-destruction of the carrier of the bump must already have been in place early in the universe.  This is the first clear detection of the bump in the host of a GRB, with the SMC being the typical type of extinction for GRB sightlines.  This makes \grb\ an ideal candidate to study the environment needed for the creation and/or non-annihilation of bump, by comparing it to other bump-less GRBs.  

To accurately derive the properties of the extinction curve in the UV and the bump we fit it to the parametrization of \citet{fitzpatrick2007}.  
The bump is found to be centered at $x_c\approx4.6\pm0.1$~$\mu$m$^{-1}$ (or $\lambda_c\approx2174\pm50$~\AA), the width of the bump is found to be $\gamma\approx1.08\pm0.05$~$\mu$m$^{-1}$ while the 'strength' is $c_3\approx2.7\pm0.1$ (taking into account the uncertainty in the intrinsic spectral slope $\beta$).  The value of $c_3$ is the same as the average that \citet{gordon2003} find for the LMC average sample ($c_3=2.7\pm0.1$) but higher than for the LMC2 sample ($c_3=1.5\pm0.1$) although it is within the scatter.  The width $\gamma$ is a bit wider compared to their $\gamma=0.93\pm0.02$~$\mu$m$^{-1}$ forethe LMC average and $\gamma=0.95\pm0.03$ for LMC2 (although it is still within the scatter of both samples).  The amount of extinction is $A_V\approx1.3$, but when taking into account $1~\sigma$ deviations in $\beta$, $A_V=0.8$--$1.5$ for the FM and LMC fits.

In the Local Group, the MW bump is a characteristic feature of the MW extinction curve.  It is also observed in the LMC, although it is usually weaker and followed by a steeper rise in the UV, while it is not observed in four out of five curves measured for the SMC (see \ref{sec:ext_ext} for more details and exceptions).  Of these three 'local type extinction laws' we find that the extinction of \grb\ most closely resembles that of the LMC.  We find that this result is robust, even taking into account a possible contribution to the extinction from the strong foreground \ion{Mg}{2} absorber.  It has been suggested, based on the difference in the SMC, LMC and MW, that the strength of the bump correlates with metallicity.  However, \citet{gordon97} found that although starburst galaxies can have varying metallicities, their extinction curves all lack the 2175~\AA\ bump.  By comparing \grb\ to another high metallicity GRB sightline which does not show any sign of a bump in its extinction curve, we similarly conclude that metallicity is not the only driver of the 2175~\AA\ bump.

Another special feature in the spectrum of \grb\ is the detection of a strong \ion{C}{1} absorption.  This is to our knowledge the first GRB spectrum to contain \ion{C}{1} absorption and we propose that there may be a correlation between its detection and the presence of the bump.  We have checked this suggested correlation for a few other lines of sight, and found all of them to be in agreement.  This prediction has also been checked by \citet{2009ApJ...691L..27P} who find a  \ion{C}{1} absorption line and a  2175~\AA\ bump (based on photometric data) in GRB~080607.
We also find a high dust-to-gas ratio, which is consistent with a proposed correlation by \citet{gordon2003}, suggesting that the strength of the bump is related to the dust-to-gas ratio.  Extending their correlation plot to include the MW, \grb\ and other GRBs from the literature with extinction analysis, we find that they all follow the proposed correlation.  Finally, the presence of the \ion{C}{1} absorption feature in GRB\,070802 suggests that the UV radiation field is weaker than in typical GRB environments.  This is in agreement with the tentative conclusion of \citet{gordon2003} that a weaker UV flux density is found to correlate with the presence of the bump.  

The simultaneous presence of \ion{C}{1} and the
2175\,\AA\ bump as well as the lower overall ionization state of the gas,
relatively (though not exceptionally) high metallicity, and large
dust-to-gas ratio may be explained in a scenario in which the dust column is
strongly enriched by the presence of asymptotic giant branch (AGB) stars.  This would be consistent with the conclusion that the host of \grb\ is a massive evolved galaxy, and supports the conclusions of \citet{noll07} that the presence of the bump requires an evolved population.

\acknowledgments
We thank Javier Gorosabel for providing us with his code for the Pei parametrization of the SMC, LMC and MW extinction laws.  We also thank P\'all Jakobsson (Palli) and the anonymous referee for their comments on the manuscript.  We thank the various members of the GRB community for regularly sacrificing their beauty sleep in their chase for GRB afterglows.  
The Dark Cosmology Centre is supported by the DNRF.  \'A.~E.  and PMV acknowledge the support of the EU under a Marie Curie International Outgoing Fellowship, contract PIOF-GA-2008-220049, and a Marie Curie Intra-European Fellowship, contract MEIF-CT-2006-041363.  J.~X.~P. is partially supported by NASA/Swift grants NNG06GJ07G and NNX07AE94G and an NSF CAREER grant (AST-0548180).

\bibliography{references}

\appendix
\section{Parametric extinction laws}
\label{app:ext}
In this Appendix we describe the parametrizations we have used to model the extinction curves.

\subsection{The Pei parametrization for the SMC and the LMC}
This parametrization was introduced by \citet{pei1992} and is given by:
\begin{eqnarray}
A_\lambda&=&A_B \sum_{i=1}^{6}\frac{a_i}{(\lambda/\lambda_i)^{n_i}+(\lambda_i/\lambda)^{n_i}+b_i}
\end{eqnarray}
where the six terms are all positive and represent different parts of the extinction curve.  The parameters can be found in \citet{pei1992}.  The five terms with $n_i=2$ are equivalent to Drude profiles with a peak at $\lambda_i$.  The only free variable in the fit is the overall amount of extinction.  The original \citet{pei1992} paper scales it to $A_B$ but we choose to scale it to $A_V$ to be consistent with the other parametrizations.  Note that the \citet{pei1992} law can also be used to describe Milky Way type of extinction.

\begin{figure*}
\begin{center}
\includegraphics[angle=0,width=0.5\columnwidth,clip=]{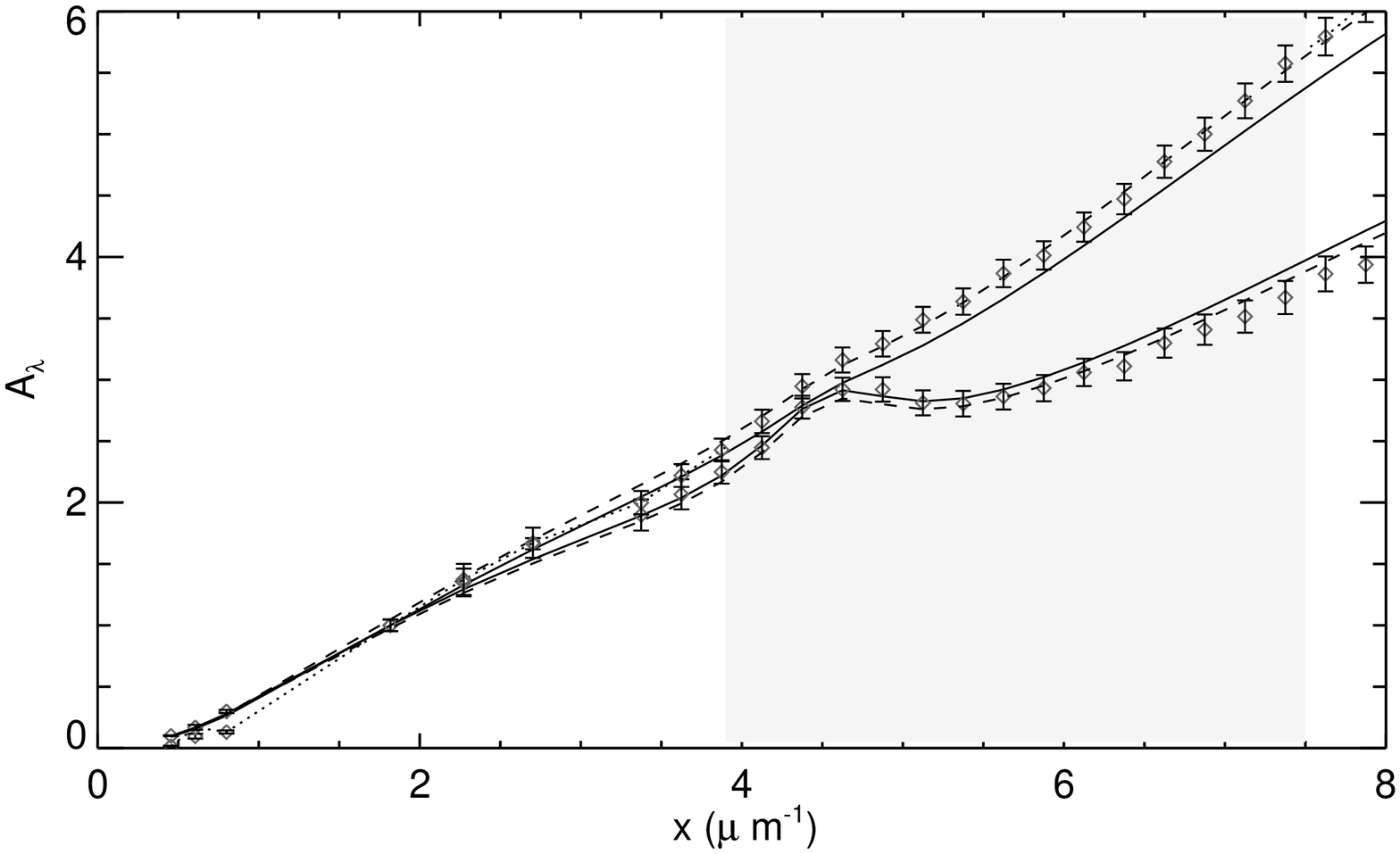}
\end{center}
\caption{A comparison of the SMC and LMC 'typical' extinction curves from \citet{pei1992} and \citet{gordon2003}.  The solid curves are the \citet{pei1992} curves for $A_V=1$, the diamonds are the \citet{gordon2003} curves for $A_V=1$ for their "LMC2" and "SMC Bar" average with error bars, and the dashed curves are \citet{pei1992} fits to the \citet{gordon2003} curves (where $A_V$ is varied).  The fits are done for points in the shaded region which is the same region as for the fits in the paper.   The derived $A_V$ differ by 5\% for the SMC curve and by 2\% for the LMC curve which is much smaller than the uncertainty in $A_V$ in our fits from the uncertainty in the intrinsic slope $\beta$.   In addition, the UV slope of the \citet{pei1992} LMC type of extinction is slightly steeper than that of \citet{gordon2003}, but falls within the 1~$\sigma$ limit.  We therefore conclude that the choice of parametrization of the SMC and LMC type of extinction does not affect our comparison of the extinction curve of \grb\ to the local type of extinction laws.}
\label{fig:app_gordon_pei}
\end{figure*}

\citet{gordon2003} present new and updated average extinction curves for the SMC and LMC extinction curves.  Their analysis is based on using the \citet{1990ApJS...72..163F} parametrization which differs from the updated FM parametrization we use (see below) in keeping $c_5$ fixed.  The \citet{gordon2003} analysis presents a more nuanced picture of the extinction in the SMC and LMC with lines of sight showing different type of extinction (see \ref{sec:ext_ext}).  However, we find that their "LMC2" average extinction curve and their "SMC Bar" extinction curves are very similar to the LMC and SMC extinction curves of \citet[][see Fig. \ref{fig:app_gordon_pei}]{pei1992}, thus justifying our use of the commonly used \citet{pei1992} parametrization.

\subsection{The CCM parametrization for Milky Way type of extinction}
This parametrization of the Milky Way extinction law was proposed by \citet{cardelli1989}.  It depends on only two parameters, $E(B-V)=A(B)-A(V)$ and $R_V=A(V)/E(B-V)$ which is the ratio of total to selective extinction.  It is given by
\begin{eqnarray}
\label{eq:car}
A_\lambda& = & E(B-V) \left[R_V a(x) + b(x)\right] \\
          & = &  A(V) \left[ a(x) + \frac{1}{R_V} b(x)\right],\nonumber
\end{eqnarray}
where $A(\lambda)$ is the total extinction at wavelength $\lambda$, $a(x)$ and $b(x)$ are polynomials and $x=\lambda^{-1}$.  The advantage of this parametrization over the one of \citet{pei1992} for the Milky Way is that it allows for a varying $R_V$.

\subsection{The FM parametrization}
The parametrisation for the UV (i.e. valid for $x>3.7$~$\mu$m$^{-1}$) is given by
\begin{eqnarray}
\label{eq:fm2007}
A_\lambda&=&E(B-V)\left(k\left(\lambda-V\right)+R_V\right)\\
& = & A(V)\left(\frac{1}{R_V}k\left(\lambda-V\right)+1\right)
\end{eqnarray}
where 
\begin{eqnarray}
\label{eq:fm2007_k}
k(\lambda-V) = 
\left\{ 
\begin{array}{ll}
c_1 + c_2 x + c_3 D(x,x_c,\gamma)                  &  x \leq c_5     \\
c_1 + c_2 x + c_3 D(x,x_c,\gamma) + c_4 (x-c_5)^2  &  x > c_5  \;\; ,
\end{array}
\right. 
\end{eqnarray}
and 
\begin{equation}
\label{eqnDRUDE}
D(x,x_c,\gamma) = \frac{x^2}{(x^2-x_c^2)^2 +x^2\gamma^2}.
\end{equation}
The  parameters $c_1$ and $c_2$ define the linear component underlying the entire UV range, $c_3$, $x_c$ and $\gamma$ give the $2175$ \AA\ bump (although its central wavelength is not fixed in the parametrization) and $c_4$ and $c_5$ give a far-UV curvature component.  The extinction properties in the infrared and optical are not parametrized in the FM2007 description but are derived using spline interpolation \citep[see ][ for details]{fitzpatrick2007}.  As our dataset does not reach into these regions in the restframe, we do not constrain the extinction curve in this region.  Therefore, for display purposes, we have chosen to set these parameters to 'typical' values found by \citet{fitzpatrick2007}, to create 'normal' smooth continuation of the curve towards $x=0$.  We note that the choice of these parameters does in no way affect our fits for the UV parameters and, vice versa, that the FM fits do not constrain this part of the curve.

As explained in \citet{fitzpatrick2007}, additional useful quantities can be defined using the UV parameters.  The first one (1) $\Delta 1250 = c_4 (8.0-c_5)^2$ gives the value of the far-UV curvature term at $1250$~\AA\ and measures the strength of the far-UV curvature; (2) $\Delta_{\mathrm{bump}}=\pi c_3/(2\gamma)$  is the area of the bump and (3) $E_{\mathrm{bump}}=c_3/\gamma^2$ is the maximum height of the bump above the linear extinction.

\end{document}